%% file: Utschick_et_al_LISA_for_Hybrid_Precoding_arxiv_version.tex
\documentclass[12pt, draftclsnofoot, onecolumn]{IEEEtran}
\usepackage[T1]{fontenc}
\usepackage{cite}
\ifCLASSINFOpdf
\else
\fi
\usepackage{url}
\usepackage{amsbsy}
\usepackage{amsmath}
\usepackage{fixmath}
\usepackage{amssymb}
\usepackage{ulem}
\usepackage[caption=false,font=footnotesize]{subfig}

\usepackage[noend]{algorithmic}
\usepackage{algorithm}

\newlength\myindent
\algsetup{indent=\myindent}
\algsetup{linenodelimiter=}

\usepackage{tikz}
\usetikzlibrary{arrows,decorations,backgrounds,shadows,plotmarks}

\usepackage{pgfplots}
\pgfplotsset{compat=newest}
\pgfplotsset{plot coordinates/math parser=false}
\pgfplotsset{every axis/.append style={font=\footnotesize}}
\newlength\figureheight
\newlength\figurewidth
\usepackage{3dplot}

\definecolor{TUMblue}{rgb}{0,0.396,.741}
\definecolor{NWSred}{rgb}{0.6,0,0}
\definecolor{TUMgray}{rgb}{.9,.9,.9}
\definecolor{Mygreen}{rgb}{0,.7,0}

\DeclareMathOperator*{\h}{H}
\DeclareMathOperator*{\T}{T}

\DeclareMathOperator*{\argmax}{argmax}
\DeclareMathOperator*{\Diag}{diag}

\DeclareMathOperator{\E}{E}

\DeclareMathOperator*{\trace}{tr}

\newcommand{\revised}[1]{\textcolor{black}{#1}}

\newcommand\copyrighttext{%
  \footnotesize \textcopyright This work has been submitted to the IEEE for possible publication. Copyright may be transferred without notice, after which this version may no longer be accessible.}
\newcommand\copyrightnotice{%
\begin{tikzpicture}[remember picture,overlay]
\node[anchor=south,yshift=10pt] at (current page.south) {\fbox{\parbox{\dimexpr\textwidth-\fboxsep-\fboxrule\relax}{\copyrighttext}}};
\end{tikzpicture}%
}

\begin{document}
\title{Hybrid LISA Precoding for Multiuser Millimeter-Wave Communications}
\author{Wolfgang~Utschick,~\IEEEmembership{Senior~Member,~IEEE,}
Christoph~St\"ockle, \IEEEmembership{Student~Member,~IEEE,} Michael~Joham,~\IEEEmembership{Member,~IEEE,} and~Jian~Luo\thanks{The authors are with Methods of Signal Processing, Technische Universit\"at M\"unchen, 80290 Munich, Germany (e-mail: \{utschick, christoph.stoeckle, joham\}@tum.de) and with Huawei Technologies D{\"u}sseldorf GmbH, Munich Office, European Research Center, (e-mail: jianluo@huawei.com).}
}
\markboth{}{}

\maketitle

\copyrightnotice
\begin{abstract}
Millimeter-wave (mmWave) 
communications plays an important role for future cellular networks because of the vast amount of spectrum available in the underutilized mmWave frequency bands. To overcome the huge free space omnidirectional path loss in those frequency bands, the deployment of a very large number of antenna elements at the base station 
is crucial. The complexity, power consumption and costs resulting from the large number of antenna elements can be reduced by limiting the number of RF chains. This leads to hybrid precoding and combining, which, in contrast to the traditional fully digital precoding and combining, moves a part of the signal processing from the digital to the analog domain. 
This paper proposes new algorithms for the design of hybrid precoders and combiners in 
a multiuser scenario. The algorithms are based on the previously proposed \textit{Linear Successive Allocation} 
method developed for the traditional fully digital version. 
It successively allocates data streams to users 
and suppresses the respective interstream interference in two stages, which perfectly matches the hybrid architecture. 
Furthermore, a low-complexity version 
is developed by exploiting the typical structure of mmWave channels. The good performance of the proposed method and its low-complexity version is demonstrated by simulation results.
\end{abstract}


\IEEEpeerreviewmaketitle

\section{Introduction}
\IEEEPARstart{T}{he} vast amount of spectrum available in the underutilized mmWave frequency bands is considered as one of the key enablers for the demanded tremendous increase in the capacity of future cellular networks. Therefore, mmWave communications plays an important role for future cellular networks \cite{Rangan2014,Roh2014,Andrews2014,Rappaport2013,Pi2011,Rappaport2014}. One of the main challenges of exploiting the spectrum in the mmWave frequency bands is the high free space omnidirectional path loss in those frequency bands \cite{Rangan2014,Roh2014,Andrews2014,Pi2011,Rappaport2014}. This problem can be overcome by the deployment of a very large number of antenna elements at the base station (BS) known as massive MIMO \cite{Marzetta2010,Rusek2013,Larsson2014,Adhikary2014}, which leads to large antenna gains \cite{Roh2014,Larsson2014}. Due to those large antenna gains, more energy can be transmitted and received through narrower directed beams, which can compensate for the high free space omnidirectional path loss \cite{Roh2014}. Therefore, massive MIMO makes the communication in the underutilized mmWave frequency bands viable and thus can increase the amount of usable spectrum. In addition, the large antenna gains used for beamforming makes massive MIMO one of the most promising methods for increasing the spectral efficiency of future cellular networks \cite{Larsson2014,NSN2013}. However, the main drawbacks of massive MIMO are the high complexity, power consumption, and costs resulting from the large number of antenna elements. 

Traditional precoding at the BS is performed digitally in the baseband. After the digital signal processing in the baseband, which modifies both the amplitude and the phase of the complex-valued data symbols, the processed signals are passed through RF chains consisting of digital-to-analog converters, mixers and power amplifiers to obtain the RF signals that are transmitted by the BS antenna elements at the carrier frequency \cite{Liang2014,Liang2014a}. This, however, requires a dedicated RF chain for each of the many BS antenna elements. Therefore, one possibility of reducing the complexity, power consumption and costs resulting from the large number of antenna elements is reducing the number of RF chains, which can be connected to the BS antenna elements 
via a network of phase shifters \cite{Liang2014a,Ayach2014,Alkhateeb2014}. This leads to hybrid precoding, where a part of the signal processing at the BS is still performed in the digital domain at baseband in front of the RF chains and the other part in the analog domain by the network of phase shifters between the RF chains and the BS antenna elements at the carrier frequency. As a consequence, the hybrid precoder consists of a digital precoder and an analog precoder, which has constant-modulus entries since we assume a fully-connected phase shifter network and only the phase can be modified by the phase shifters. In the design of hybrid precoders this special structure has to be taken into account.

Several works deal with the design of hybrid precoders for single-user mmWave systems. In \cite{Ayach2014}, e.g., the optimal fully digital precoder is approximated by a precoder that consists of an analog and a digital precoder. By exploiting the spatially sparse structure of the mmWave channels and restricting the columns of the analog precoder to be from a dictionary of array response vectors naturally having constant-modulus entries, the approximation by means of hybrid precoding 
is formulated as a sparse recovery problem, which is solved by an algorithm based on the sparse recovery \textit{Orthogonal Matching Pursuit} (OMP) method from compressed sensing (CS). 
In order to avoid the restriction to the dictionary of array response vectors and the high computational complexity in case of high-resolution dictionaries, a dictionary-free algorithm for approximating the optimal fully digital precoder by the hybrid precoder is proposed in \cite{Mendez-Rial2015}. 
Assuming that the optimal equalizer is used by the MS, the authors of \cite{Sohrabi2016} suggest to iteratively determine the analog precoder for the BS and, given the already designed analog precoder, to determine subsequently the optimal digital precoder. The analog precoder is determined iteratively by sequentially updating each element of the analog precoder while keeping all other elements fixed. A similar procedure is applied afterwards to design the analog and digital equalizer for the MS.

Due to the large number of antenna elements, the BS can serve several MSs in the same time-frequency resource by spatial multiplexing \cite{Larsson2014}. Therefore, hybrid precoding solutions are urgently required for such multiuser scenarios.  In \cite{Sohrabi2016}, an algorithm for the design of hybrid precoders in a multiuser MISO scenario, where the BS serves several single-antenna MSs, is proposed in addition to that for the single-user case mentioned before. Similarly to the single-user case, the analog precoder is determined iteratively for a fixed digital precoder. For a fixed analog precoder, the digital precoder is determined by zero-forcing, which suppresses the multiuser interference, and a power allocation such that the power constraint is fulfilled. The alternation between designing the analog and the digital units eventually results in a multiuser hybrid precoder. By contrast, the authors of \cite{Liang2014a} suggest to simply match the analog precoder to the channel from the BS to the single-antenna MSs by normalizing all elements of its Hermitian to the same modulus and keeping only their phases, and to suppress the occurring multiuser interference by the digital precoder performing low-dimensional zero-forcing precoding. For a multiuser scenario, where each MS has one RF chain but possibly several antenna elements, a two-stage algorithm designing the hybrid precoder for the BS and the analog equalizers for the MSs is presented in \cite{Alkhateeb2015a} and \cite{Alkhateeb2015}. In the first stage of this algorithm, called \textit{Two-Stage Multi-User Hybrid Precoders Algorithm} (2SMUHPA), the desired signal power for each MS is maximized by choosing the columns of the analog precoder and the analog equalizers from sets of array response vectors while neglecting the resulting multiuser interference, which is suppressed in the second stage by the digital precoder. 

In this paper, we propose a new multiuser hybrid precoding method for the general setting, where each MS receives an arbitrary number of data streams up to the number of its RF chains. The new scheme is based on the \textit{Linear Successive Allocation} (LISA) method previously proposed for the fully digital precoding in multiuser MIMO systems (see \cite{Guthy2008, Guthy2009,GuUtHuJo10}). The LISA method is a linear version of the earlier proposed \textit{Successive Encoding and Successive Allocation Method} (SESAM) \cite{TeUtBaNo05}, \cite{Tejera2006}, that combines the successive allocation of data streams to MSs with a successive encoding technique based on the coding technique with known interference introduced in \cite{Costa1983}. Its excellent properties have been demonstrated in measurement campaigns \cite{BaBaAnGuHeNiNoTeUt07,BaTeGuUtNoHeBaAnNiStKh07} and later confirmed by means of large system analysis results in \cite{GuUtHo10b,GuUtHo13}. LISA successively allocates data streams to the MSs, and determines the precoders and equalizers for those data streams, which circumvents the high computational complexity of the direct sum rate maximization while maintaining the performance of state-of-the-art methods for achieving high sum rates in multiuser MIMO systems. 

\revised{\textit{Our contributions are as follows:}} 

\revised{(1) We show that the previously proposed LISA method ideally matches the requirements of the hybrid precoding architecture. In contrast to all state-of-the-art techniques, where the decomposition into an analog and digital part must be enforced by an artificial decomposition of the fully digital precoding solutions or by iterative design rules of the respective parts, the proposed method includes the decomposition step as an inherent feature.} 

\revised{(2) By a simple elementwise normalization of an intermediate solution of LISA to obtain the analog precoder (first stage) and the construction of a successive digital precoder (second stage) to suppress the remaining interference, the new Hybrid LISA (H-LISA) method for multiuser hybrid precoding is proposed. 
}

\revised{(3) Furthermore, by exploiting the typical geometric nature of the channel matrices for mmWave channels, we present a new low-complexity version of the proposed methods that clearly outperforms state-of-the-art methods in terms of achievable sum rate, while keeping the numerical complexity at a similar level. 
}

The paper is organized as follows. In Sec. \ref{sec2}, the considered system model is introduced. After reviewing the 2SMUHPA, Sec. \ref{sec3} describes how the hybrid version can be obtained from LISA, while Sec.~\ref{sec4} deals with reducing the computational complexity of both, LISA and H-LISA. In Sec.~\ref{sec5}, H-LISA is modified for analog processing at the MSs. Numerical results for the comparison of the proposed solutions are presented in Sec.~\ref{sec6} and Sec.~\ref{sec7} finally concludes the paper.

We use the following notation throughout this paper: The absolute value and the phase of a complex valued scalar $a$ are written as $\left|a\right|$ and $\arg\left(a\right)$, respectively. Bold lower and upper case letters denote vectors and matrices, respectively. The $\left\|\cdot\right\|_2$ is the Euclidean norm of a vector, while $\left(\cdot\right)^{\h}$ represents the Hermitian of a vector or a matrix, and $\left[\cdot\right]_{i,j}$, $\trace{\left(\cdot\right)}$, and $\det\left(\cdot\right)$ are the element in the $i$-th row and the $j$-th column of a matrix, its trace, and determinant, respectively. The $\Diag\left(\cdot\right)$ operator sets all off-diagonal elements of a matrix to $0$, whereas $\Diag\left(a_1,a_2,\ldots,a_N\right)$ is the diagonal matrix whose diagonal elements are $a_1,a_2,\ldots,a_N$. The $N$-dimensional identity matrix is written as $\mathbf{I}_N$, and $\operatorname{span}\left\{\cdot\right\}$, $\operatorname{null}\left\{\cdot\right\}$, and $\left(\cdot\right)^{\perp}$ denote the span, the nullspace, and the orthogonal complement, respectively. $\boldsymbol{a}\sim\mathcal{CN}\left(\boldsymbol{m},\boldsymbol{R}\right)$ is a vector following the circularly symmetric complex multivariate Gaussian distribution with mean $\boldsymbol{m}$ and covariance matrix $\boldsymbol{R}$, and $\E\left[\cdot\right]$ is the expectation.

\section{System Model}\label{sec2}

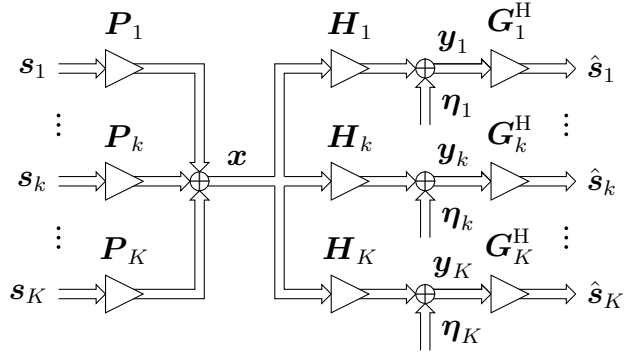
\begin{figure}
\centering

\begin{tikzpicture}[scale=0.25]


\begin{scope}[yshift=6cm]
\draw (-12,0) node[left]{$\boldsymbol{s}_1$};
\draw (-12,0.25) -- ++(2,0) -- ++(0,0.25) -- ++(0.5,-0.5);
\draw (-12,-0.25) -- ++(2,0) -- ++(0,-0.25) -- ++(0.5,0.5);

\draw (-8.5,-0.25) -- ++(3.75,0) -- ++(0,-4.75) -- ++(-0.25,0) -- ++(0.5,-0.5);
\draw (-8.5,0.25) -- ++(4.25,0)-- ++(0,-5.25) -- ++(0.25,0) -- ++(-0.5,-0.5);

\draw[fill=white] (-9.5,-1) -- ++(0,2) -- ++(2,-1) -- cycle;
\draw (-9.5,-1) ++(1,2) node[above]{$\boldsymbol{P}_1$};

\draw (0,-5.75) -- ++(0,5.5) -- ++(2,0) -- ++(0,-0.25) -- ++(0.5,0.5);
\draw (-0.5,-5.75) -- ++(0,6) -- ++(2.5,0) -- ++(0,0.25) -- ++(0.5,-0.5);

\draw (3.5,0.25) -- ++(3,0) -- ++(0,0.25) -- ++(0.5,-0.5);
\draw (3.5,-0.25) -- ++(3,0) -- ++(0,-0.25) -- ++(0.5,0.5);

\draw[fill=white] (2.5,-1) -- ++(0,2) -- ++(2,-1) -- cycle;
\draw (2.5,-1) ++(1,2) node[above]{$\boldsymbol{H}_1$};

\draw (7.25,-3) -- ++(0,2) -- ++(-0.25,0) -- ++(0.5,0.5);
\draw (7.75,-3) -- ++(0,2) node[midway,right]{$\boldsymbol{\eta}_1$} -- ++(0.25,0) -- ++(-0.5,0.5);

\draw (7.5,0.25) -- ++(3,0) -- ++(0,0.25) -- ++(0.5,-0.5);
\draw (7.5,0.25) -- ++(3,0) node[midway,above]{$\boldsymbol{y}_1$} -- ++(0,0.25) -- ++(0.5,-0.5);
\draw (7.5,-0.25) -- ++(3,0) -- ++(0,-0.25) -- ++(0.5,0.5);

\draw (12,0.25) -- ++(3,0) -- ++(0,0.25) -- ++(0.5,-0.5) node[right]{$\hat{\boldsymbol{s}}_1$};
\draw (12,-0.25) -- ++(3,0) -- ++(0,-0.25) -- ++(0.5,0.5);

\draw[fill=white] (11,-1) -- ++(0,2) -- ++(2,-1) -- cycle;
\draw (11,-1) ++(1,2) node[above]{$\boldsymbol{G}_1^{\h}$};



\draw[fill=white] (7.5,0) node{$+$} circle(0.5);
\end{scope}

\begin{scope}
\draw (-12,0) node[left]{$\boldsymbol{s}_k$};
\draw (-12,0.25) -- ++(2,0) -- ++(0,0.25) -- ++(0.5,-0.5);
\draw (-12,-0.25) -- ++(2,0) -- ++(0,-0.25) -- ++(0.5,0.5);

\fill[black] (-12,3.5) circle(0.1) ++(0,-0.5) circle(0.1) ++(0,-0.5) circle(0.1);
\fill[black] (-12,-2.5) circle(0.1) ++(0,-0.5) circle(0.1) ++(0,-0.5) circle(0.1);

\draw (-8.5,0.25) -- ++(3,0) -- ++(0,0.25) -- ++(0.5,-0.5);
\draw (-8.5,-0.25) -- ++(3,0) -- ++(0,-0.25) -- ++(0.5,0.5);

\draw[fill=white] (-9.5,-1) -- ++(0,2) -- ++(2,-1) -- cycle;
\draw (-9.5,-1) ++(1,2) node[above]{$\boldsymbol{P}_k$};

\draw (0,0.25) -- ++(2,0) -- ++(0,0.25) -- ++(0.5,-0.5);
\draw (0,-0.25) -- ++(2,0) -- ++(0,-0.25) -- ++(0.5,0.5);

\draw (3.5,0.25) -- ++(3,0) -- ++(0,0.25) -- ++(0.5,-0.5);
\draw (3.5,-0.25) -- ++(3,0) -- ++(0,-0.25) -- ++(0.5,0.5);

\draw[fill=white] (2.5,-1) -- ++(0,2) -- ++(2,-1) -- cycle;
\draw (2.5,-1) ++(1,2) node[above]{$\boldsymbol{H}_k$};

\draw (7.25,-3) -- ++(0,2) -- ++(-0.25,0) -- ++(0.5,0.5);
\draw (7.75,-3) -- ++(0,2) node[midway,right]{$\boldsymbol{\eta}_k$} -- ++(0.25,0) -- ++(-0.5,0.5);

\draw (7.5,0.25) -- ++(3,0) node[midway,above]{$\boldsymbol{y}_k$} -- ++(0,0.25) -- ++(0.5,-0.5);
\draw (7.5,0.25) -- ++(3,0) -- ++(0,0.25) -- ++(0.5,-0.5);
\draw (7.5,-0.25) -- ++(3,0) -- ++(0,-0.25) -- ++(0.5,0.5);

\draw (12,0.25) -- ++(3,0) -- ++(0,0.25) -- ++(0.5,-0.5) node[right]{$\hat{\boldsymbol{s}}_k$};
\draw (12,-0.25) -- ++(3,0) -- ++(0,-0.25) -- ++(0.5,0.5);

\draw[fill=white] (11,-1) -- ++(0,2) -- ++(2,-1) -- cycle;
\draw (11,-1) ++(1,2) node[above]{$\boldsymbol{G}_k^{\h}$};

\fill[black] (15,3.5) circle(0.1) ++(0,-0.5) circle(0.1) ++(0,-0.5) circle(0.1);
\fill[black] (15,-2.5) circle(0.1) ++(0,-0.5) circle(0.1) ++(0,-0.5) circle(0.1);



\draw[fill=white] (7.5,0) node{$+$} circle(0.5);

\draw (-4.5,0.25) -- node[midway,above]{$\boldsymbol{x}$} ++(4,0);
\draw (-4.5,-0.25) -- ++(4,0);
\draw[fill=white] (-4.5,0) node{$+$} circle(0.5);

\end{scope}

\begin{scope}[yshift=-6cm]
\draw (-12,0) node[left]{$\boldsymbol{s}_K$};
\draw (-12,0.25) -- ++(2,0) -- ++(0,0.25) -- ++(0.5,-0.5);
\draw (-12,-0.25) -- ++(2,0) -- ++(0,-0.25) -- ++(0.5,0.5);

\draw (-8.5,0.25) -- ++(3.75,0) -- ++(0,4.75) -- ++(-0.25,0) -- ++(0.5,0.5);
\draw (-8.5,-0.25) -- ++(4.25,0)-- ++(0,5.25) -- ++(0.25,0) -- ++(-0.5,0.5);

\draw[fill=white] (-9.5,-1) -- ++(0,2) -- ++(2,-1) -- cycle;
\draw (-9.5,-1) ++(1,2) node[above]{$\boldsymbol{P}_K$};

\draw (0,5.75) -- ++(0,-5.5) -- ++(2,0) -- ++(0,0.25) -- ++(0.5,-0.5);
\draw (-0.5,5.75) -- ++(0,-6) -- ++(2.5,0) -- ++(0,-0.25) -- ++(0.5,0.5);

\draw (3.5,0.25) -- ++(3,0) -- ++(0,0.25) -- ++(0.5,-0.5);
\draw (3.5,-0.25) -- ++(3,0) -- ++(0,-0.25) -- ++(0.5,0.5);

\draw[fill=white] (2.5,-1) -- ++(0,2) -- ++(2,-1) -- cycle;
\draw (2.5,-1) ++(1,2) node[above]{$\boldsymbol{H}_K$};

\draw (7.25,-3) -- ++(0,2) -- ++(-0.25,0) -- ++(0.5,0.5);
\draw (7.75,-3) -- ++(0,2) node[midway,right]{$\boldsymbol{\eta}_K$} -- ++(0.25,0) -- ++(-0.5,0.5);

\draw (7.5,0.25) -- ++(3,0) node[midway,above]{$\boldsymbol{y}_K$} -- ++(0,0.25) -- ++(0.5,-0.5);
\draw (7.5,0.25) -- ++(3,0) -- ++(0,0.25) -- ++(0.5,-0.5);
\draw (7.5,-0.25) -- ++(3,0) -- ++(0,-0.25) -- ++(0.5,0.5);

\draw (12,0.25) -- ++(3,0) -- ++(0,0.25) -- ++(0.5,-0.5) node[right]{$\hat{\boldsymbol{s}}_K$};
\draw (12,-0.25) -- ++(3,0) -- ++(0,-0.25) -- ++(0.5,0.5);

\draw[fill=white] (11,-1) -- ++(0,2) -- ++(2,-1) -- cycle;
\draw (11,-1) ++(1,2) node[above]{$\boldsymbol{G}_K^{\h}$};



\draw[fill=white] (7.5,0) node{$+$} circle(0.5);
\end{scope}

\end{tikzpicture}
\caption{System model.}
\label{fig:system_model}
\end{figure}

In the downlink, the BS equipped with $N_\text{BS}$ transmit antenna elements communicates to $K$ MSs $k\in\left\{1,2,\ldots,K\right\}$, each of which has $N_{\text{MS}} \leq N_{\text{BS}}$ receive antenna elements. As depicted in Fig.~\ref{fig:system_model}, the BS forms the transmitted signal vector $ \boldsymbol{x} = \sum_{k=1}^{K}\boldsymbol{P}_k\boldsymbol{s}_k\in\mathbb{C}^{N_\text{BS}} $ 
from the vectors $\boldsymbol{s}_k\in\mathbb{C}^{d_k}$ consisting of the data symbols to be transmitted to the MSs by using the precoders $\boldsymbol{P}_k\in\mathbb{C}^{N_\text{BS} \times d_k}$.
The $d_k$ elements of $\boldsymbol{s}_k\sim\mathcal{CN}\left(\boldsymbol{0},\mathbf{I}_{d_k}\right)$ are the data symbols intended for the $k^\text{th}$ MS, where $d_k \leq N_{\text{MS}}$ is the number of its data streams.
Since hybrid precoding is applied at the BS with $N_\text{RF}<N_\text{BS}$ RF chains, the total number of data streams $d=\sum_{k=1}^{K}d_k \leq N_\text{RF}$ is limited by the number of RF chains and the precoder $\boldsymbol{P}_k$ has the special structure 
\begin{equation}
\boldsymbol{P}_k=\boldsymbol{P}_{\text{A}}\boldsymbol{P}_{\text{D},k},
\label{eq:P_k}
\end{equation}
where $\boldsymbol{P}_{\text{D},k}\in\mathbb{C}^{N_\text{RF} \times d_k}$ is the digital precoder for the $k^{\text{th}}$ MS and $\boldsymbol{P}_{\text{A}}\in\mathbb{C}^{N_\text{BS} \times N_\text{RF}}$ is the analog precoder implemented by phase shifters. Each phase shifter connects one of the $N_\text{RF}$ RF chains with one of the $N_\text{BS}$ BS antenna elements and allows only adjustments of the phase. As a consequence, the analog precoder is the same for all $K$ MSs and is restricted to have constant-modulus entries \cite{Alkhateeb2015}. More specifically, $\boldsymbol{P}_{\text{A}}\in\mathcal{P}^{N_\text{BS} \times N_\text{RF}}$ with $\mathcal{P}=\left\{p\in\mathbb{C}:\left|p\right|=\frac{1}{\sqrt{N_\text{BS}}}\right\}$. Since only the total average transmit power $P$ is available, the precoders $\boldsymbol{P}_k$ have to fulfill the total power constraint $ \E\left[\left\|\boldsymbol{x}\right\|_2^2\right]=\sum_{k=1}^{K}\trace\left(\boldsymbol{P}_k\boldsymbol{P}_k^{\h}\right) \leq P $.

Adopting a narrowband block-fading channel model, the received signal vector $\boldsymbol{y}_k\in\mathbb{C}^{N_\text{MS}}$ of the $k^{\text{th}}$ MS reads as $ \boldsymbol{y}_{k}=\boldsymbol{H}_k\boldsymbol{x}+\boldsymbol{\eta}_k $,
where the channel matrix $\boldsymbol{H}_k\in\mathbb{C}^{N_\text{MS} \times N_\text{BS}}$ characterizes the channel between the BS and the $k^{\text{th}}$ MS, and the noise vector $\boldsymbol{\eta}_k\sim\mathcal{CN}\left(\boldsymbol{0},\mathbf{I}_{N_\text{MS}}\right)$ reflects the noise corrupting the received signal.
Due to the high free space omnidirectional path loss and signal attenuation in the mmWave frequency bands, the scattering is limited \cite{Rangan2014} such that there might be only a small number of paths over which the signals from the BS can reach the $k^{\text{th}}$ MS. This allows to use a geometric channel model, where the channel matrix is of the special form \cite{Alkhateeb2015}
\begin{equation}
\begin{split}
\boldsymbol{H}_k = \sqrt{\frac{N_\text{BS}N_\text{MS}}{L_k}}
\sum_{\ell=1}^{L_k}\alpha_{k,\ell}\boldsymbol{a}_\text{MS}\left(\phi^\text{MS}_{k,\ell},\theta^\text{MS}_{k,\ell}\right)\boldsymbol{a}_\text{BS}^{\h}\left(\phi^\text{BS}_{k,\ell},\theta^\text{BS}_{k,\ell}\right).
\end{split}
\label{eq:H_k}
\end{equation}
Here, $L_k$ denotes the number of paths between the BS and the $k^{\text{th}}$ MS while the $\ell^{\text{th}}$ path is characterized by the complex path gain $\alpha_{k,\ell}\sim\mathcal{CN}\left(0,1\right)$, the azimuth and elevation angle of departure (AoD) $\phi^\text{BS}_{k,\ell}$ and $\theta^\text{BS}_{k,\ell}$ at the BS as well as the azimuth and elevation angle of arrival (AoA) $\phi^\text{MS}_{k,\ell}$ and $\theta^\text{MS}_{k,\ell}$ at the MS. The vectors $\boldsymbol{a}_\text{MS}(\phi^\text{MS}_{k,\ell},\theta^\text{MS}_{k,\ell})$ and $\boldsymbol{a}_\text{BS}(\phi^\text{BS}_{k,\ell},\theta^\text{BS}_{k,\ell})$ are the array response vectors of the antenna arrays deployed at the MS and BS, respectively. \revised{For an interelement spacing of half the carrier wavelength, the azimuth angle $\phi$ and the elevation angle $\theta$, the array response vector of an 
$M \times N $ uniform planar array (UPA) with totally $ MN \in \{N_{\text{BS}},N_\text{{MS}}\} $ antenna elements is equal to
$ \boldsymbol{a}_\text{UPA}\left(\phi,\theta\right) = \frac{1}{\sqrt{MN}}\left[1,\ldots,\text{e}^{j\pi\left(m\sin\phi\sin\theta+n\cos\theta\right)},\ldots,\text{e}^{j\pi\left(\left(M-1\right)\sin\phi\sin\theta+\left(N-1\right)\cos\theta\right)}\right]^{\T} $
with $m\in\left\{0,1,\ldots,M-1\right\}$ and $n\in\left\{0,1,\ldots,N-1\right\}$ \cite{Ayach2014}.}

By applying the equalizer $\boldsymbol{G}_k\in\mathbb{C}^{N_\text{MS} \times d_k}$ to the received signal vector $\boldsymbol{y}_k$, the $k^{\text{th}}$ MS obtains the 
signal
\begin{equation}
\hat{\boldsymbol{s}}_k
=\boldsymbol{G}_k^{\h}\boldsymbol{H}_k\sum_{j=1}^{K}\boldsymbol{P}_j\boldsymbol{s}_j+\boldsymbol{G}_k^{\h}\boldsymbol{\eta}_k.
\label{eq:system_model}
\end{equation}

\section{Determination of Precoders and Equalizers}\label{sec3}

\begin{figure}[h!]
\centering
\begin{tikzpicture}[scale=0.25]
%
\fill[red!50] (-4,0) -- (0,-4) -- (0,8) -- (-4,4);
\fill[red!50] (0,-4) -- (4,-2) -- (4,6) -- (0,8);
\draw[latex reversed-latex reversed] (-4,0) -- (-4,4) node[midway,left]{$N_\text{RF}$};
\draw[latex reversed-latex reversed] (0,-4) -- (0,8) node[midway,left]{$N_\text{BS}$};
\draw[latex reversed-latex reversed] (4,-2) -- (4,6) node[midway,right]{$KN_\text{MS}$};
\end{tikzpicture}
\caption{Illustration of the bottleneck of the system.}
\label{fig:bottleneck}
\end{figure}
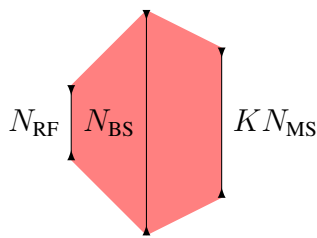

We would like to determine the hybrid precoders $\boldsymbol{P}_k$ with the special structure in (\ref{eq:P_k}) and the equalizers $\boldsymbol{G}_k$ such that the sum rate
\begin{equation}
R_\text{sum}=\sum_{k=1}^K\log_2\frac{\det\left(\boldsymbol{G}_k^{\h}\boldsymbol{G}_k+\boldsymbol{G}_k^{\h}\boldsymbol{H}_k\sum\limits_{j=1}^K\boldsymbol{P}_j\boldsymbol{P}_j^{\h}\boldsymbol{H}_k^{\h}\boldsymbol{G}_k\right)}{\det\left(\boldsymbol{G}_k^{\h}\boldsymbol{G}_k+\boldsymbol{G}_k^{\h}\boldsymbol{H}_k\sum\limits_{\substack{j=1\\j \neq k}}^K\boldsymbol{P}_j\boldsymbol{P}_j^{\h}\boldsymbol{H}_k^{\h}\boldsymbol{G}_k\right)}
\label{eq:sum_rate}
\end{equation}
of all $K$ MSs becomes maximum and the total average power constraint is fulfilled. So, we aim at solving the constrained optimization problem
\begin{equation}
\begin{split}
\max_{\left\{\boldsymbol{P}_k,\boldsymbol{G}_k\right\}_{k=1}^{K}}
R_\text{sum} \quad
\text{s.t.} \quad & \sum_{k=1}^{K}\trace\left(\boldsymbol{P}_k\boldsymbol{P}_k^{\h}\right) \leq P, \enspace \boldsymbol{P}_k=\boldsymbol{P}_{\text{A}}\boldsymbol{P}_{\text{D},k}, \\ 
& \boldsymbol{P}_{\text{A}}\in\mathcal{P}^{N_\text{BS} \times N_\text{RF}}, \enspace \boldsymbol{P}_{\text{D},k}\in\mathbb{C}^{N_\text{RF} \times d_k}, \enspace \boldsymbol{G}_k\in\mathbb{C}^{N_\text{MS} \times d_k} \enspace \forall k.
\end{split}
\label{eq:sum_rate_maximization}
\end{equation}
If the total number of receive antenna elements $KN_\text{MS}$ is larger than the number of RF chains $N_\text{RF}$, which limits the total number of data streams $d$, not all MSs can receive the maximum number of data streams $N_\text{MS}$. So, the number of RF chains $N_\text{RF}$ can be expected to be the bottleneck of the system as illustrated in Fig.~\ref{fig:bottleneck}.

The optimal allocation of the data streams to the MSs, which determines the number of data streams $d_k$ for the individual MSs, becomes a combinatorial problem involving an exhaustive search. Even if the data stream allocation is fixed, the direct solution of (\ref{eq:sum_rate_maximization}) is computationally intractable because of the non-concavity of the objective function and the non-convex constraints on the analog precoder resulting from the phase shifters.

\subsection{Two-Stage Multi-User Hybrid Precoders Algorithm (2SMUHPA)}

The algorithm proposed in \cite{Alkhateeb2015a} and \cite{Alkhateeb2015} circumvents the direct solution of (\ref{eq:sum_rate_maximization}) by following a heuristic approach and is restricted to the special case, where there is exactly one data stream per MS and RF chain such that $N_\text{RF}=K=d$. In this case, the input and output vectors of the system model, $\boldsymbol{s}_k$ and $\hat{\boldsymbol{s}}_k$, become scalars $s_k$ and $\hat{s}_k$, respectively, while the precoders $\boldsymbol{P}_k$ reduce to vectors $\boldsymbol{p}_k=\boldsymbol{P}_{\text{A}}\boldsymbol{p}_{\text{D},k}\in\mathbb{C}^{N_\text{BS}}$ consisting of an analog precoding matrix $\boldsymbol{P}_{\text{A}}$ and a digital precoding vector $\boldsymbol{p}_{\text{D},k}\in\mathbb{C}^{N_\text{RF}}$, and the equalizers $\boldsymbol{G}_k$ to vectors $\boldsymbol{g}_k\in\mathbb{C}^{N_\text{MS}}$. As a consequence, the processed received symbol for the data symbol $s_k$ of the $k^{\text{th}}$ data stream intended for the $k^{\text{th}}$ MS can be written as
\begin{equation}
\hat{s}_k
=\boldsymbol{g}_k^{\h}\boldsymbol{H}_k\boldsymbol{P}_{\text{A}}\boldsymbol{p}_{\text{D},k}s_k
+\sum\limits_{\substack{j=1\\j \neq k}}^K\boldsymbol{g}_k^{\h}\boldsymbol{H}_k\boldsymbol{P}_{\text{A}}\boldsymbol{p}_{\text{D},j}s_j
+\boldsymbol{g}_k^{\h}\boldsymbol{\eta}_k,
\label{eq:system_model_2SMUHPA}
\end{equation}
which follows from (\ref{eq:system_model}).
In addition to $\boldsymbol{P}_{\text{A}}=\left[\boldsymbol{p}_{\text{A},1},\boldsymbol{p}_{\text{A},2},\ldots,\boldsymbol{p}_{\text{A},K}\right]$, the equalizers $\boldsymbol{g}_k$ shall be implemented by phase shifters in the analog domain, too. In order to fulfill the constraints resulting from the analog processing, the column $\boldsymbol{p}_{\text{A},k}$ of the analog precoder $\boldsymbol{P}_{\text{A}}$ and the equalizer $\boldsymbol{g}_k$ are selected from beamsteering codebooks $\mathcal{P}_{\text{A},k}$ consisting of array response vectors of the form $\boldsymbol{a}_\text{BS}\left(\phi,\theta\right)$ and $\mathcal{G}_k$ consisting of array response vectors of the form $\boldsymbol{a}_\text{MS}\left(\phi,\theta\right)$, respectively. An overview of the algorithm, which is called Two-Stage Multi-User Hybrid Precoders Algorithm (2SMUHPA), is given in Algorithm~\ref{algo:2SMUHPA}.
\begin{algorithm}
\caption{Two-Stage Multi-User Hybrid Precoders Algorithm (2SMUHPA)}
\label{algo:2SMUHPA}
\begin{algorithmic}[1]
    \REQUIRE $\left\{\boldsymbol{H}_k\right\}_{k=1}^{K}$, $P$, $\left\{\mathcal{G}_k\right\}_{k=1}^{K}$, $\left\{\mathcal{P}_{\text{A},k}\right\}_{k=1}^{K}$
        \STATE \textbf{Stage 1:}
		        \renewcommand{\algorithmicfor}{\hspace{\myindent}\textbf{for}}
		        \renewcommand{\algorithmicendfor}{\hspace{\myindent}\algorithmicend\ \textbf{for}}
				\FOR{$k=1,2,\ldots,K$}
				\STATE \hspace{\myindent}$\displaystyle\left\{\boldsymbol{g}_k,\boldsymbol{p}_{\text{A},k}\right\}=\argmax_{\boldsymbol{g}\in\mathcal{G}_k,\boldsymbol{p}_{\text{A}}\in\mathcal{P}_{\text{A},k}}
\left|\boldsymbol{g}^{\h}\boldsymbol{H}_k\boldsymbol{p}_{\text{A}}\right|$ \label{algo:2SMUHPA:line:maximization}
				\ENDFOR
				\STATE \hspace{\myindent}$\boldsymbol{P}_{\text{A}}=\left[\boldsymbol{p}_{\text{A},1},\boldsymbol{p}_{\text{A},2},\ldots,\boldsymbol{p}_{\text{A},K}\right]$
        \STATE \textbf{Stage 2:}
				\STATE \hspace{\myindent}$\hat{\boldsymbol{H}}=
\begin{bmatrix}
\boldsymbol{g}_1^{\h}\boldsymbol{H}_1\boldsymbol{P}_{\text{A}}\\
\boldsymbol{g}_2^{\h}\boldsymbol{H}_2\boldsymbol{P}_{\text{A}}\\
\vdots\\
\boldsymbol{g}_K^{\h}\boldsymbol{H}_K\boldsymbol{P}_{\text{A}}
\end{bmatrix}$, $\boldsymbol{\Gamma}=\frac{P}{K}\mathbf{I}_K$, $\boldsymbol{\Lambda}=\left[\Diag\left(\left(\boldsymbol{P}_{\text{A}}\hat{\boldsymbol{H}}^{-1}\right)^{\h}\boldsymbol{P}_{\text{A}}\hat{\boldsymbol{H}}^{-1}\right)\right]^{-\frac{1}{2}}$
				\STATE \hspace{\myindent}$\boldsymbol{P}_{\text{D}}=\left[\boldsymbol{p}_{\text{D},1},\boldsymbol{p}_{\text{D},2},\ldots,\boldsymbol{p}_{\text{D},K}\right]=\hat{\boldsymbol{H}}^{-1}\boldsymbol{\Lambda}\boldsymbol{\Gamma}^{\frac{1}{2}}$
		        \renewcommand{\algorithmicfor}{\textbf{for}}
		        \renewcommand{\algorithmicendfor}{\algorithmicend\ \textbf{for}}
			  \FOR{$k=1,2,\ldots,K$}
				\STATE $\boldsymbol{P}_k=\boldsymbol{P}_{\text{A}}\boldsymbol{p}_{\text{D},k}$, $\boldsymbol{G}_k=\boldsymbol{g}_k$
				\ENDFOR
    \ENSURE $\left\{\boldsymbol{P}_k\right\}_{k=1}^{K}$, $\left\{\boldsymbol{G}_k\right\}_{k=1}^{K}$
\end{algorithmic}
\end{algorithm}

The algorithm, which is called Two-Stage Multi-User Hybrid Precoders Algorithm (2SMUHPA) consists of two stages. In the first stage, the desired signal power for each user $k$ is maximized while neglecting the multiuser or interstream interference $\sum_{j=1,j \neq k}^K\boldsymbol{g}_k^{\h}\boldsymbol{H}_k\boldsymbol{P}_{\text{A}}\boldsymbol{p}_{\text{D},j}s_j$ in (\ref{eq:system_model_2SMUHPA}) to obtain the analog equalizers and columns of the analog precoder $\boldsymbol{P}_{\text{A}}$.
The column $\boldsymbol{p}_{\text{A},k}$ of the analog precoder $\boldsymbol{P}_{\text{A}}$ and the analog equalizer $\boldsymbol{g}_k$ for the MS $k$ are chosen to be the array response vectors $\boldsymbol{a}_\text{BS}(\phi^\text{BS}_{k,\ell},\theta^\text{BS}_{k,\ell})$ and $\boldsymbol{a}_\text{MS}(\phi^\text{MS}_{k,\ell},\theta^\text{MS}_{k,\ell})$ corresponding to the strongest path between the BS and the MS $k$ with the largest magnitude of the complex path gain $\alpha_{k,\ell}$ in order to try to maximize the desired signal power for the MS $k$.
In the second stage, the interference neglected so far is suppressed by exploiting the remaining degrees of freedom, namely, the digital precoders.
The digital precoder $\boldsymbol{p}_{\text{D},j}$ of the $j^{\text{th}}$ data stream for the $j^{\text{th}}$ MS must lie in the nullspace of the effective channels $\boldsymbol{g}_k^{\h}\boldsymbol{H}_k\boldsymbol{P}_{\text{A}}$ of all other data streams, i.e., $\boldsymbol{p}_{\text{D},j}\in\operatorname{null}\left\{\boldsymbol{g}_k^{\h}\boldsymbol{H}_k\boldsymbol{P}_{\text{A}}\right\}$, $j \neq k$, such that the interference is suppressed, i.e., $\sum_{j=1,j \neq k}^K\boldsymbol{g}_k^{\h}\boldsymbol{H}_k\boldsymbol{P}_{\text{A}}\boldsymbol{p}_{\text{D},j}s_j=0$.
To this end, all effective channels are collected in the composite channel matrix
\begin{equation}
\hat{\boldsymbol{H}}=
\begin{bmatrix}
\boldsymbol{g}_1^{\h}\boldsymbol{H}_1\boldsymbol{P}_{\text{A}}\\
\boldsymbol{g}_2^{\h}\boldsymbol{H}_2\boldsymbol{P}_{\text{A}}\\
\vdots\\
\boldsymbol{g}_K^{\h}\boldsymbol{H}_K\boldsymbol{P}_{\text{A}}
\end{bmatrix}
\in\mathbb{C}^{K \times K}
\end{equation}
and the digital precoder $\boldsymbol{P}_{\text{D}}=\left[\boldsymbol{p}_{\text{D},1},\boldsymbol{p}_{\text{D},2},\ldots,\boldsymbol{p}_{\text{D},K}\right]$, whose columns are the individual digital precoders $\boldsymbol{p}_{\text{D},k}$, is determined by
\begin{equation}
\boldsymbol{P}_{\text{D}}=\hat{\boldsymbol{H}}^{-1}\boldsymbol{\Lambda}\boldsymbol{\Gamma}^{\frac{1}{2}}.
\end{equation}
Here, the diagonal matrix $ \boldsymbol{\Lambda}=\left[\Diag\left(\left(\boldsymbol{P}_{\text{A}}\hat{\boldsymbol{H}}^{-1}\right)^{\h}\boldsymbol{P}_{\text{A}}\hat{\boldsymbol{H}}^{-1}\right)\right]^{-\frac{1}{2}} $
normalizes the columns of $\boldsymbol{P}_{\text{A}}\hat{\boldsymbol{H}}^{-1}$ to unit norm and $\boldsymbol{\Gamma}$ is a diagonal power loading matrix.
As a consequence, 
the composite channel is diagonalized and decomposes to $K$ scalar interference-free subchannels, whose channel gains are the diagonal elements of $\boldsymbol{\Lambda}$, whereas the diagonal elements of $\boldsymbol{\Gamma}$ are the powers given to those scalar subchannels. In \cite{Alkhateeb2015a} and \cite{Alkhateeb2015}, the available total average transmit power $P$ is distributed equally among the $K$ subchannels and equal power $P/K$ is given to all of them such that $ \boldsymbol{\Gamma}=\frac{P}{K}\mathbf{I}_K $.
Although not considered in \cite{Alkhateeb2015a} and \cite{Alkhateeb2015}, the optimal powers can be determined by waterfilling. Waterfilling might allocate zero power to the scalar subchannel of a MS, which effectively does not get a data stream then. As a consequence, each MS gets at most $1$ data stream in contrast to the original version of the 2SMUHPA without waterfilling, where each MS receives exactly $1$ data stream.
Finally, $\boldsymbol{p}_k=\boldsymbol{P}_{\text{A}}\boldsymbol{p}_{\text{D},k}$ form the hybrid precoders $\boldsymbol{P}_k$ and $\boldsymbol{g}_k$ the analog equalizers $\boldsymbol{G}_k$.

\subsection{Linear Successive Allocation (LISA)}

Linear Successive Allocation (LISA) previously developed for the traditional fully digital precoding is an algorithm that circumvents the exhaustive search required if not all MSs can receive the maximum number of data streams $N_\text{MS}$. \revised{It avoids the high computational complexity of directly solving the non-convex sum-rate maximization problem (\ref{eq:sum_rate_maximization}) by successively allocating data streams to the MSs and determining the precoders and equalizers for those data streams respectively \cite{Guthy2009}. In contrast to the 2SMUHPA, it is not restricted to the special case of at most $1$ data stream per MS. Similarly to the 2SMUHPA, it is also based on two stages for suppressing the interstream interference solely by means of linear signal processing, which perfectly matches the requirements for the analog and digital part of the hybrid precoding architecture.} Therefore, we propose to use LISA for hybrid precoding and adapt it to the constraints resulting from the analog processing. \revised{In the following, the main characteristics of LISA are described.}

\begin{figure}
\centering

\begin{tikzpicture}[scale=0.25]


\begin{scope}[yshift=6cm]
\draw (-12,0) node[left]{$t_1$};
\draw[-latex] (-12,0) -- ++(2.5,0);

\draw (-8.5,-0.25) -- ++(3.75,0) -- ++(0,-4.75) -- ++(-0.25,0) -- ++(0.5,-0.5);
\draw (-8.5,0.25) -- ++(4.25,0)-- ++(0,-5.25) -- ++(0.25,0) -- ++(-0.5,-0.5);

\draw[fill=white] (-9.5,-1) -- ++(0,2) -- ++(2,-1) -- cycle;
\draw (-9.5,-1) ++(1,2) node[above]{$\boldsymbol{p}_1$};

\draw (0,-5.75) -- ++(0,5.5) -- ++(2,0) -- ++(0,-0.25) -- ++(0.5,0.5);
\draw (-0.5,-5.75) -- ++(0,6) -- ++(2.5,0) -- ++(0,0.25) -- ++(0.5,-0.5);

\draw (3.5,0.25) -- ++(3,0) -- ++(0,0.25) -- ++(0.5,-0.5);
\draw (3.5,-0.25) -- ++(3,0) -- ++(0,-0.25) -- ++(0.5,0.5);

\draw[fill=white] (2.5,-1) -- ++(0,2) -- ++(2,-1) -- cycle;
\draw (2.5,-1) ++(1,2) node[above]{$\boldsymbol{H}_{\pi\left(1\right)}$};

\draw (7.25,-3) -- ++(0,2) -- ++(-0.25,0) -- ++(0.5,0.5);
\draw (7.75,-3) -- ++(0,2) node[midway,right]{$\boldsymbol{\eta}_{\pi\left(1\right)}$} -- ++(0.25,0) -- ++(-0.5,0.5);

\draw (7.5,0.25) -- ++(3,0) -- ++(0,0.25) -- ++(0.5,-0.5);
\draw (7.5,-0.25) -- ++(3,0) -- ++(0,-0.25) -- ++(0.5,0.5);

\draw[-latex] (12,0) -- ++(3.5,0) node[right]{$\hat{t}_1$};

\draw[fill=white] (11,-1) -- ++(0,2) -- ++(2,-1) -- cycle;
\draw (11,-1) ++(1,2) node[above]{$\boldsymbol{g}_1^{\h}$};



\draw[fill=white] (7.5,0) node{$+$} circle(0.5);
\end{scope}

\begin{scope}
\draw (-12,0) node[left]{$t_i$};
\draw[-latex] (-12,0) -- ++(2.5,0);

\fill[black] (-12,3.5) circle(0.1) ++(0,-0.5) circle(0.1) ++(0,-0.5) circle(0.1);
\fill[black] (-12,-2.5) circle(0.1) ++(0,-0.5) circle(0.1) ++(0,-0.5) circle(0.1);

\draw (-8.5,0.25) -- ++(3,0) -- ++(0,0.25) -- ++(0.5,-0.5);
\draw (-8.5,-0.25) -- ++(3,0) -- ++(0,-0.25) -- ++(0.5,0.5);

\draw[fill=white] (-9.5,-1) -- ++(0,2) -- ++(2,-1) -- cycle;
\draw (-9.5,-1) ++(1,2) node[above]{$\boldsymbol{p}_i$};

\draw (0,0.25) -- ++(2,0) -- ++(0,0.25) -- ++(0.5,-0.5);
\draw (0,-0.25) -- ++(2,0) -- ++(0,-0.25) -- ++(0.5,0.5);

\draw (3.5,0.25) -- ++(3,0) -- ++(0,0.25) -- ++(0.5,-0.5);
\draw (3.5,-0.25) -- ++(3,0) -- ++(0,-0.25) -- ++(0.5,0.5);

\draw[fill=white] (2.5,-1) -- ++(0,2) -- ++(2,-1) -- cycle;
\draw (2.5,-1) ++(1,2) node[above]{$\boldsymbol{H}_{\pi\left(i\right)}$};

\draw (7.25,-3) -- ++(0,2) -- ++(-0.25,0) -- ++(0.5,0.5);
\draw (7.75,-3) -- ++(0,2) node[midway,right]{$\boldsymbol{\eta}_{\pi\left(i\right)}$} -- ++(0.25,0) -- ++(-0.5,0.5);

\draw (7.5,0.25) -- ++(3,0) -- ++(0,0.25) -- ++(0.5,-0.5);
\draw (7.5,-0.25) -- ++(3,0) -- ++(0,-0.25) -- ++(0.5,0.5);

\draw[-latex] (12,0) -- ++(3.5,0) node[right]{$\hat{t}_i$};

\draw[fill=white] (11,-1) -- ++(0,2) -- ++(2,-1) -- cycle;
\draw (11,-1) ++(1,2) node[above]{$\boldsymbol{g}_i^{\h}$};



\draw[fill=white] (7.5,0) node{$+$} circle(0.5);

\fill[black] (15,3.5) circle(0.1) ++(0,-0.5) circle(0.1) ++(0,-0.5) circle(0.1);
\fill[black] (15,-2.5) circle(0.1) ++(0,-0.5) circle(0.1) ++(0,-0.5) circle(0.1);

\draw (-4.5,0.25) -- node[midway,above]{$\boldsymbol{x}$} ++(4,0);
\draw (-4.5,-0.25) -- ++(4,0);
\draw[fill=white] (-4.5,0) node{$+$} circle(0.5);

\end{scope}

\begin{scope}[yshift=-6cm]

\draw (-12,0) node[left]{$t_d$};
\draw[-latex] (-12,0) -- ++(2.5,0);

\draw (-8.5,0.25) -- ++(3.75,0) -- ++(0,4.75) -- ++(-0.25,0) -- ++(0.5,0.5);
\draw (-8.5,-0.25) -- ++(4.25,0)-- ++(0,5.25) -- ++(0.25,0) -- ++(-0.5,0.5);

\draw[fill=white] (-9.5,-1) -- ++(0,2) -- ++(2,-1) -- cycle;
\draw (-9.5,-1) ++(1,2) node[above]{$\boldsymbol{p}_d$};

\draw (0,5.75) -- ++(0,-5.5) -- ++(2,0) -- ++(0,0.25) -- ++(0.5,-0.5);
\draw (-0.5,5.75) -- ++(0,-6) -- ++(2.5,0) -- ++(0,-0.25) -- ++(0.5,0.5);

\draw (3.5,0.25) -- ++(3,0) -- ++(0,0.25) -- ++(0.5,-0.5);
\draw (3.5,-0.25) -- ++(3,0) -- ++(0,-0.25) -- ++(0.5,0.5);

\draw[fill=white] (2.5,-1) -- ++(0,2) -- ++(2,-1) -- cycle;
\draw (2.5,-1) ++(1,2) node[above]{$\boldsymbol{H}_{\pi\left(d\right)}$};

\draw (7.25,-3) -- ++(0,2) -- ++(-0.25,0) -- ++(0.5,0.5);
\draw (7.75,-3) -- ++(0,2) node[midway,right]{$\boldsymbol{\eta}_{\pi\left(d\right)}$} -- ++(0.25,0) -- ++(-0.5,0.5);

\draw (7.5,0.25) -- ++(3,0) -- ++(0,0.25) -- ++(0.5,-0.5);
\draw (7.5,-0.25) -- ++(3,0) -- ++(0,-0.25) -- ++(0.5,0.5);

\draw[-latex] (12,0) -- ++(3.5,0) node[right]{$\hat{t}_d$};

\draw[fill=white] (11,-1) -- ++(0,2) -- ++(2,-1) -- cycle;
\draw (11,-1) ++(1,2) node[above]{$\boldsymbol{g}_d^{\h}$};



\draw[fill=white] (7.5,0) node{$+$} circle(0.5);
\end{scope}

\end{tikzpicture}
\caption{System model for LISA.}
\label{fig:system_model_2}
\end{figure}
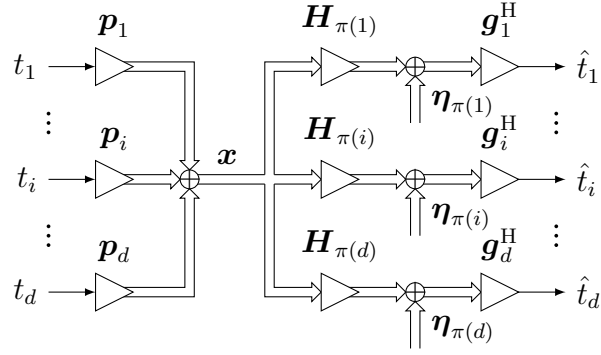

The function $ \pi: \left\{1,2,\ldots,d\right\} \rightarrow \left\{1,2,\ldots,K\right\}, i \mapsto \pi\left(i\right) $
keeps track of the successive allocation of the data streams to the MSs, i.e., $\pi\left(i\right)$ indicates to which MS the $i^{\text{th}}$ data stream is allocated and $d_{k,i}$ denotes the number of data streams that have been allocated to the MS $k$ after the $i^{\text{th}}$ allocation step. \revised{Therefore, the data symbol $t_i$ of the $i^{\text{th}}$ data stream is an element 
of the input signal vector $\boldsymbol{s}_{\pi\left(i\right)}$ and the corresponding precoder $\boldsymbol{p}_i$ of the $i^{\text{th}}$ data stream forms a column 
of the precoder $\boldsymbol{P}_{\pi\left(i\right)}$ 
(cf. Fig.~\ref{fig:system_model}). Furthermore, the processed received symbol $\hat{t}_i$ 
of the $i^{\text{th}}$ data stream is an element 
of the output signal vector $\hat{\boldsymbol{s}}_{\pi\left(i\right)}$ of the system model and the corresponding equalizer $\boldsymbol{g}_i$ of the $i^{\text{th}}$ data stream forms a column 
of the equalizer $\boldsymbol{G}_{\pi\left(i\right)}$. 
With this, it follows from (\ref{eq:system_model}) that $ \hat{t}_i = \boldsymbol{g}_i^{\h}\boldsymbol{H}_{\pi\left(i\right)}\sum_{j=1}^{d}\boldsymbol{p}_jt_j+\boldsymbol{g}_i^{\h}\boldsymbol{\eta}_{\pi\left(i\right)} $.
This converts the representation of the system model 
in Fig.~\ref{fig:system_model} to the alternative 
scalar representation in Fig.~\ref{fig:system_model_2}, where the inputs and outputs refer to the $ d = \sum_{k=1}^{K}d_k $ data symbols of the individual data streams.
}
Similarly to (\ref{eq:system_model_2SMUHPA}), the processed received symbol for the data symbol $t_i$ can be written as
\begin{equation}
\hat{t}_i = \boldsymbol{g}_i^{\h}\boldsymbol{H}_{\pi\left(i\right)}\boldsymbol{p}_it_i
+\sum_{\substack{j=1\\j \neq i}}^{d}\boldsymbol{g}_i^{\h}\boldsymbol{H}_{\pi\left(i\right)}\boldsymbol{p}_jt_j
+\boldsymbol{g}_i^{\h}\boldsymbol{\eta}_{\pi\left(i\right)}.
\label{eq:t_i_hat}
\end{equation}
For the $i^\text{th}$ data stream, the interference 
from all other data streams $j \neq i$ can be split into the interference 
from the previously allocated data streams $j < i$ and the interference 
from the successively allocated data streams $j > i$. 
In order to suppress any interstream interference, the precoder $\boldsymbol{p}_j$ of the $j^\text{th}$ data stream must lie in the nullspace of the effective channels $\boldsymbol{g}_i^{\h}\boldsymbol{H}_{\pi\left(i\right)}$ of all other data streams, i.e., $\boldsymbol{p}_{j}\in\operatorname{null}\left\{\boldsymbol{g}_i^{\h}\boldsymbol{H}_{\pi\left(i\right)}\right\}$, $j \neq i$. 

\begin{figure}
\centering

\begin{tikzpicture}[scale=0.25]

\begin{scope}[yshift=6cm]
\draw (-12,0) node[left]{$t_1$};
\draw[-latex] (-12,0) -- (-6,0);

\draw[fill=white] (-6,-1) -- ++(0,2) -- ++(2,-1) -- cycle;
\draw (-6,-1) ++(1,2) node[above]{$\boldsymbol{g}_1^{\h}\boldsymbol{H}_{\pi\left(1\right)}\boldsymbol{p}_1$};

\draw[-latex] (-4,0) -- (1.5,0);

\draw[-latex] (2,-3) -- ++(0,2.5) node[midway,right]{$\boldsymbol{g}_1^{\h}\boldsymbol{\eta}_{\pi\left(1\right)}$};

\draw[-latex] (2,0) -- (10,0) node[right]{$\hat{t}_1$};

\draw[fill=white] (2,0) node{$+$} circle(0.5);
\end{scope}

\begin{scope}
\draw (-12,0) node[left]{$t_i$};
\draw[-latex] (-12,0) -- (-6,0);

\fill[black] (-12,3.5) circle(0.1) ++(0,-0.5) circle(0.1) ++(0,-0.5) circle(0.1);
\fill[black] (-12,-2.5) circle(0.1) ++(0,-0.5) circle(0.1) ++(0,-0.5) circle(0.1);

\draw[fill=white] (-6,-1) -- ++(0,2) -- ++(2,-1) -- cycle;
\draw (-6,-1) ++(1,2) node[above]{$\boldsymbol{g}_i^{\h}\boldsymbol{H}_{\pi\left(i\right)}\boldsymbol{p}_i$};

\draw[-latex] (-4,0) -- (1.5,0);

\draw[-latex] (2,-3) -- ++(0,2.5) node[midway,right]{$\boldsymbol{g}_i^{\h}\boldsymbol{\eta}_{\pi\left(i\right)}$};

\draw[-latex] (2,0) -- (10,0) node[right]{$\hat{t}_i$};

\draw[fill=white] (2,0) node{$+$} circle(0.5);

\fill[black] (9.5,3.5) circle(0.1) ++(0,-0.5) circle(0.1) ++(0,-0.5) circle(0.1);
\fill[black] (9.5,-2.5) circle(0.1) ++(0,-0.5) circle(0.1) ++(0,-0.5) circle(0.1);

\end{scope}

\begin{scope}[yshift=-6cm]

\draw (-12,0) node[left]{$t_d$};
\draw[-latex] (-12,0) -- (-6,0);

\draw[fill=white] (-6,-1) -- ++(0,2) -- ++(2,-1) -- cycle;
\draw (-6,-1) ++(1,2) node[above]{$\boldsymbol{g}_d^{\h}\boldsymbol{H}_{\pi\left(d\right)}\boldsymbol{p}_d$};

\draw[-latex] (-4,0) -- (1.5,0);

\draw[-latex] (2,-3) -- ++(0,2.5) node[midway,right]{$\boldsymbol{g}_d^{\h}\boldsymbol{\eta}_{\pi\left(d\right)}$};

\draw[-latex] (2,0) -- (10,0) node[right]{$\hat{t}_d$};

\draw[fill=white] (2,0) node{$+$} circle(0.5);
\end{scope}

\end{tikzpicture}
\caption{Scalar interference-free subchannels produced by LISA.}
\label{fig:system_model_3}
\end{figure}
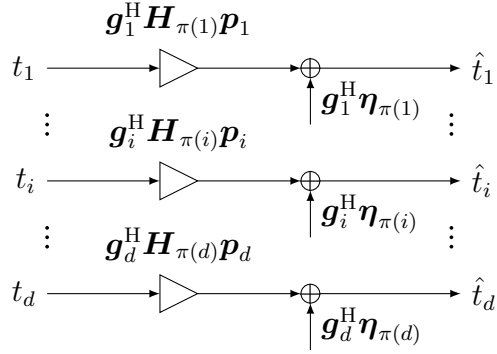

LISA takes \textit{two stages} for finding those precoders and suppressing the interference. In the \textit{first step}, it determines an \revised{auxiliary precoder} $\boldsymbol{q}_j$ for each data stream $j$ which takes into account the nullspace constraint of the effective channels of the previously allocated data streams $i<j$, i.e., $\boldsymbol{q}_{j}\in\operatorname{null}\left\{\boldsymbol{g}_i^{\h}\boldsymbol{H}_{\pi\left(i\right)}\right\}_{i=1}^{j-1}$, such that, for each data stream, \revised{only} the interference from the successively allocated data streams is suppressed. \revised{Given the \revised{auxiliary precoders} $\boldsymbol{q}_j$, LISA  determines the precoders $\boldsymbol{p}_j$ for each data stream $j$ in the \textit{second stage} of the method, such that eventually each precoder $\boldsymbol{p}_j$ lies in the nullspace of all other effective channels, i.e., 
$\boldsymbol{p}_{j}\in\operatorname{null}\left\{\boldsymbol{g}_i^{\h}\boldsymbol{H}_{\pi\left(i\right)}\right\}$, $i \neq j$, and, for each data stream, the interference from all other data streams is suppressed.} As a consequence, the expression for the processed received symbol $\hat{t}_i$ in (\ref{eq:t_i_hat}) simplifies to
\begin{equation}
\hat{t}_i = \boldsymbol{g}_i^{\h}\boldsymbol{H}_{\pi\left(i\right)}\boldsymbol{p}_it_i
+\boldsymbol{g}_i^{\h}\boldsymbol{\eta}_{\pi\left(i\right)}
\label{eq:t_i_hat_2}
\end{equation}
and the system model depicted in Fig.~\ref{fig:system_model_2} decomposes to $d$ scalar interference-free subchannels for the $d$ data streams shown in Fig.~\ref{fig:system_model_3}.
The $i^\text{th}$ 
scalar interference-free subchannel 
has the overall channel gain $\left|\boldsymbol{g}_i^{\h}\boldsymbol{H}_{\pi\left(i\right)}\boldsymbol{p}_i\right|=\sqrt{\gamma_i}\lambda_i$, which can be split into the square root of the power $\gamma_i=\left\|\boldsymbol{p}_i\right\|_2^2$ allocated to this subchannel and the actual subchannel gain \revised{$\lambda_i=\left|\boldsymbol{g}_i^{\h}\boldsymbol{H}_{\pi\left(i\right)}\boldsymbol{p}_i\right|/{\left\|\boldsymbol{p}_i\right\|_2}$}. 
As $\boldsymbol{g}_i^{\h}\boldsymbol{H}_{\pi\left(i\right)}\boldsymbol{p}_it_i\sim\mathcal{CN}\left(0,\gamma_i\lambda_i^2\right)$ and assuming $\boldsymbol{g}_i^{\h}\boldsymbol{\eta}_{\pi\left(i\right)}\sim\mathcal{CN}\left(0,1\right)$ by a respective normalization of the equalizers $\boldsymbol{g}_i$, the rate of the $i^\text{th}$ scalar subchannel is given by $ R_i=\log_2\left(1+\gamma_i\lambda_i^2\right) $.
\revised{In the following, the successive nature of LISA and its two stages are outlined in greater detail.}
\subsubsection{First Stage of LISA}
\revised{When allocating the $i$th data stream, the previously determined assignments $\pi\left(j\right)$ of data streams $j<i$  
and their respective equalizers $\boldsymbol{g}_j$ and precoders $\boldsymbol{q}_j$ 
remain fixed 
and the MS $\pi\left(i\right)$ is selected with respect to (\ref{eq:selectionrule}). The unit-norm equalizer $\boldsymbol{g}_i$ and the unit-norm precoder $\boldsymbol{q}_i$ for the $i^{\text{th}}$ data stream are chosen such that the gain of the corresponding scalar subchannel within the nullspace of the effective channels of the previously allocated data streams $j<i$, $\operatorname{null}\left\{\boldsymbol{g}_j^{\h}\boldsymbol{H}_{\pi\left(j\right)}\right\}_{j=1}^{i-1}$, is maximal, i.e.,}
\begin{equation}
\left\{\pi\left(i\right),\boldsymbol{g}_i,\boldsymbol{q}_i\right\}=
\argmax_{\substack{k\in\left\{1,2,\ldots,K\right\}\\\boldsymbol{g}\in\mathbb{C}^{N_\text{MS}},\boldsymbol{q}\in\mathbb{C}^{N_\text{BS}}}}
\left|\boldsymbol{g}^{\h}\boldsymbol{H}_k\boldsymbol{q}\right| \quad \text{s.t.} \quad 
\boldsymbol{q}\in\operatorname{null}\left\{\boldsymbol{g}_j^{\h}\boldsymbol{H}_{\pi\left(j\right)}\right\}_{j=1}^{i-1}, \quad 
\left\|\boldsymbol{g}\right\|_2=\left\|\boldsymbol{q}\right\|_2=1,
\label{eq:selectionrule}
\end{equation}
\revised{where $\operatorname{null}\left\{\boldsymbol{g}_j^{\h}\boldsymbol{H}_{\pi\left(j\right)}\right\}_{j=1}^{i-1} $ is replaced by $ \mathbb{C}^{N_\text{BS}}$ for $i=1$.}

\begin{figure}
\centering

\tdplotsetmaincoords{60}{125}

\begin{tikzpicture}[scale=0.7]
		

\tdplotsetrotatedcoords{0}{-20}{0}

\fill[Mygreen,opacity=0.5,tdplot_rotated_coords] (0,-4,-1) -- (0,4,-1) -- (0,4,0) -- (0,-4,0) -- cycle;
\fill[TUMblue,opacity=0.5,tdplot_main_coords] (-3,-4,0) -- (4,-4,0) -- (4,4,0) -- (-3,4,0) -- cycle;
\fill[Mygreen,opacity=0.5,tdplot_rotated_coords] (0,-4,0) -- (0,4,0) -- (0,4,5) -- (0,-4,5) -- cycle;


\draw[black,thick,->,tdplot_main_coords] (0,0,0) -- ++(0,0,4) node[above]{$\boldsymbol{H}_{\pi\left(1\right)}^{\h}\boldsymbol{g}_1$};
\draw[red,thick,->,tdplot_main_coords] (0,0,0) -- ++(0,0,2) node[left]{$\boldsymbol{q}_1$};
\draw[TUMblue,tdplot_main_coords] (5,5,0) node{$\operatorname{null}\left\{\boldsymbol{g}_1^{\h}\boldsymbol{H}_{\pi\left(1\right)}\right\}$};
\draw[black,thick,->,tdplot_main_coords] (0,0,0) -- ++(3.5,0,2) node[left]{$\boldsymbol{H}_{\pi\left(2\right)}^{\h}\boldsymbol{g}_2$};
\draw[dashed,tdplot_main_coords] (3.5,0,2) -- (3.5,0,0);
\draw[tdplot_main_coords] (3,0,0) -- ++(0,0,0.5) -- ++(0.5,0,0);
\draw[black,thick,->,tdplot_main_coords] (0,0,0) -- ++(3.5,0,0) node[left]{$\boldsymbol{T}_2\boldsymbol{H}_{\pi\left(2\right)}^{\h}\boldsymbol{g}_2$};
\draw[red,thick,->,tdplot_main_coords] (0,0,0) -- ++(2,0,0) node[below]{$\boldsymbol{q}_2$};
\draw[Mygreen,tdplot_rotated_coords] (0,2,6.2) node{$\operatorname{null}\left\{\boldsymbol{g}_2^{\h}\boldsymbol{H}_{\pi\left(2\right)}\right\}$};

\tdplotsetrotatedcoords{0}{20}{0}
\draw[black,thick,->,tdplot_rotated_coords] (0,0,0) -- ++(0,3,4) node[above]{$\boldsymbol{H}_{\pi\left(3\right)}^{\h}\boldsymbol{g}_3$};
\draw[dotted,semithick,tdplot_main_coords] (0,3,0) -- ++({4*sin(20)},0,0) -- ++(0,-3,0) ({4*sin(20)},3,0) -- ++(0,0,{4*cos(20)});
\draw[dashed,tdplot_rotated_coords] (0,3,4) -- (0,3,0);
\draw[tdplot_rotated_coords] (0,2.5,0) -- ++(0,0,0.5) -- ++(0,0.5,0);
\draw[black,thick,->,tdplot_rotated_coords] (0,0,0) -- ++(0,3,0) node[right]{$\boldsymbol{T}_3\boldsymbol{H}_{\pi\left(3\right)}^{\h}\boldsymbol{g}_3$};
\draw[red,thick,->,tdplot_main_coords] (0,0,0) -- ++(0,2,0) node[anchor=north east]{$\boldsymbol{q}_3$};

\tdplotsetthetaplanecoords{130}


\end{tikzpicture}
\caption{Illustration of how LISA determines the equalizers $\boldsymbol{g}_i$ and the precoders $\boldsymbol{q}_i$ for the allocation of $3$ data streams $i=1,2,3$.}
\label{fig:LISA}
\end{figure}
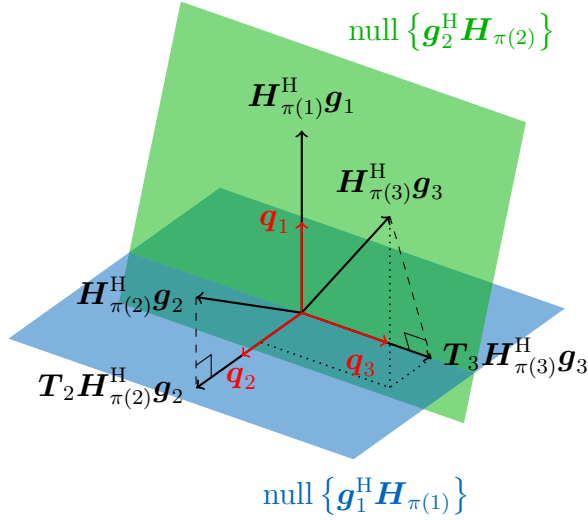

\revised{Introducing the orthogonal projector $ \boldsymbol{T}_{i+1}=\boldsymbol{T}_i-\boldsymbol{q}_i\boldsymbol{q}_i^{\h} $ onto $\operatorname{null}\left\{\boldsymbol{g}_j^{\h}\boldsymbol{H}_{\pi\left(j\right)}\right\}_{j=1}^{i}$ with $ \boldsymbol{T}_1=\mathbf{I}_{N_\text{BS}}$, the selection rule in (\ref{eq:selectionrule}) can be rewritten as
\begin{equation}
\left\{\pi\left(i\right),\boldsymbol{g}_i,\boldsymbol{q}_i\right\}=
\argmax_{\substack{k\in\left\{1,2,\ldots,K\right\}\\\boldsymbol{g}\in\mathbb{C}^{N_\text{MS}},\boldsymbol{q}\in\mathbb{C}^{N_\text{BS}}}}
\left|\boldsymbol{g}^{\h}\boldsymbol{H}_k\boldsymbol{T}_i\boldsymbol{q}\right| \quad \text{s.t.} \quad 
\left\|\boldsymbol{g}\right\|_2=\left\|\boldsymbol{q}\right\|_2=1,
\label{eq:LISA_allocation}
\end{equation}
with $
\boldsymbol{q}_i
=\left(\boldsymbol{g}^{\h}_i\boldsymbol{H}_{\pi\left(i\right)}\boldsymbol{T}_i\right)^{\h}/{\left\|\boldsymbol{g}^{\h}_i\boldsymbol{H}_{\pi\left(i\right)}\boldsymbol{T}_i\right\|_2} $. The latter obviously represents a Gram-Schmidt process, which computes orthonormal basis vectors $\left\{\boldsymbol{q}_j\right\}_{j=1}^{i}$ for the span of the vectors $\left\{\boldsymbol{H}^{\h}_{\pi\left(j\right)}\boldsymbol{g}_j\right\}_{j=1}^{i}$.
}
In essence, the MS $\pi\left(i\right)$, to which the $i^{\text{th}}$ data stream is allocated, is the MS $k$ with the largest maximum singular value  of its projected channel matrix 
and the equalizer $\boldsymbol{g}_i$ and the precoder $\boldsymbol{q}_i$ are the corresponding left and right singular vectors.
Hence, the maximal value of the objective is the maximum singular value $ \sigma_{\text{max},i} $ 
of $\boldsymbol{H}_{\pi\left(i\right)}\boldsymbol{T}_i$. 
\revised{The successive steps of LISA are illustrated in Fig.~\ref{fig:LISA} for the allocation of three data streams $i=1,2,3$.}

\revised{In a further interpretation of the first stage of LISA, the auxiliary precoders $\boldsymbol{q}_j$ of the first $i$ data streams $j=1,2,\ldots,i$ can be interpreted as elements of the orthogonal factor of the LQ-decomposition of the 
composite channel matrix, i.e., $ \boldsymbol{H}_{\text{comp},i}\boldsymbol{Q}_i=\boldsymbol{L}_i $ with
\begin{equation}
\boldsymbol{H}_{\text{comp},i}=
\begin{bmatrix}
\boldsymbol{g}_1^{\h}\boldsymbol{H}_{\pi\left(1\right)} \\
\boldsymbol{g}_2^{\h}\boldsymbol{H}_{\pi\left(2\right)} \\
\vdots \\
\boldsymbol{g}_i^{\h}\boldsymbol{H}_{\pi\left(i\right)}
\end{bmatrix}
\in\mathbb{C}^{i \times N_\text{BS}}
\quad \text{and} \quad
\boldsymbol{Q}_{i}=
\begin{bmatrix}
\boldsymbol{q}_1 &
\boldsymbol{q}_2 &
\ldots &
\boldsymbol{q}_i
\end{bmatrix}\in\mathbb{C}^{N_\text{BS} \times i}.
\label{eq:Q_i}
\end{equation}
}

After the completion of the first step, the product of the composite channel matrix $\boldsymbol{H}_{\text{comp},d}$ containing the effective channels $\boldsymbol{g}_i^{\h}\boldsymbol{H}_{\pi\left(i\right)}$ of all $d$ data streams $i=1,2,\ldots,d$ and the matrix $\boldsymbol{Q}_d$ containing the precoders $\boldsymbol{q}_{j}$ of all $d$ data streams $j=1,2,\ldots,d$ is a lower triangular matrix $ \boldsymbol{H}_{\text{comp},d}\boldsymbol{Q}_d=\boldsymbol{L}_d $
such that the elements $\boldsymbol{g}_i^{\h}\boldsymbol{H}_{\pi\left(i\right)}\boldsymbol{q}_j$, $j>i$, above the main diagonal are $0$ and, for each data stream $i$, only the interference from the successively allocated data streams $j>i$ is suppressed.

\subsubsection{Second Stage of LISA} \revised{The nested structure of nullspaces generated during the first step of LISA, which subsequent precoders must be element of, inherently guarantees that data streams which are allocated later do not interfere with earlier allocated data streams. On the other hand, due to the greedy nature of the method, earlier assigned precoders cannot take into account their interference to later allocated data streams. Consequently,} in order to suppress also the remaining interference from the previously allocated data streams $j<i$ for each data stream $i$, 
the effective precoder
\begin{equation}
\boldsymbol{P}_{\text{eff},i}=
\begin{bmatrix}
\boldsymbol{p}_1 &
\boldsymbol{p}_2 &
\ldots &
\boldsymbol{p}_i
\end{bmatrix}
\in\mathbb{C}^{N_\text{BS} \times i},
\end{equation}
whose columns are the precoders $\boldsymbol{p}_{j}$ of the first $i$ data streams $j=1,2,\ldots,i$, is determined as
\begin{equation}
\boldsymbol{P}_{\text{eff},i}=\boldsymbol{Q}_i\boldsymbol{L}_i^{-1}\boldsymbol{\Lambda}_i\boldsymbol{\Gamma}_i^{\frac{1}{2}}.
\label{eq:P_eff_i}
\end{equation}
Here, the diagonal matrix \revised{$ \boldsymbol{\Lambda}_i 
=\Diag\left(\lambda_{i,1},\lambda_{i,2},\ldots,\lambda_{i,i}\right)
=\left[\Diag\left(\left(\boldsymbol{L}_i^{-1}\right)^{\h}\boldsymbol{L}_i^{-1}\right)\right]^{-\frac{1}{2}} $}
normalizes the columns of $\boldsymbol{Q}_i\boldsymbol{L}_i^{-1}$ to unit norm and \revised{$\boldsymbol{\Gamma}_i=\Diag\left(\gamma_{i,1},\gamma_{i,2},\ldots,\gamma_{i,i}\right)$} is a diagonal power loading matrix.
This choice of the effective precoder $\boldsymbol{P}_{\text{eff},i}$ ensures that the product of the composite channel matrix $\boldsymbol{H}_{\text{comp},i}$ and $\boldsymbol{P}_{\text{eff},i}$ is the diagonal matrix
\begin{equation}
\boldsymbol{H}_{\text{comp},i}\boldsymbol{P}_{\text{eff},i}=\boldsymbol{\Lambda}_i\boldsymbol{\Gamma}_i^{\frac{1}{2}}.
\end{equation}
The rate of the $j^\text{th}$ scalar subchannel after allocating the $i^\text{th}$ data stream is given by $ R_{i,j}=\log_2\left(1+\gamma_{i,j}\lambda_{i,j}^2\right) $ 
and the sum rate after allocating the $i^\text{th}$ data stream reads $ R_{\text{sum},i}=\sum_{j=1}^{i}R_{i,j} $.
Given the subchannel gains $\left\{\lambda_{i,j}\right\}_{j=1}^{i}$, the optimal power values
\begin{equation}
\left\{\gamma_{i,j}\right\}_{j=1}^{i}
=\argmax_{\left\{\gamma_{j}\right\}_{j=1}^{i}} \sum_{j=1}^{i}\log_2\left(1+\gamma_j\lambda_{i,j}^2\right) 
\quad \text{s.t.} \quad \sum_{j=1}^{i}\gamma_{j} \leq P,\quad \gamma_{j}\geq0 \enspace \forall j,
\end{equation}
which are maximizing the sum rate $R_{\text{sum},i}$ while fulfilling the power constraint, 
are determined by waterfilling. \revised{In \cite{Guthy2009}, it is shown that the whole procedure of selecting MSs, equalizers and precoders maximizes a lower bound for the sum rate $R_{\text{sum},i}$.}


\revised{Increasing the number of allocated data streams obviously imposes more zero-forcing constraints to be taken into account in the second stage of LISA. Those additional zero-forcing constraints might lead to a decrease in the channel gains of the eventually scalar interference-free subchannels, which in turn might even reduce the sum rate. Therefore, the successive allocation of data streams is only continued, if allocating  a further data stream actually increases the overall sum rate, i.e., $R_{\text{sum},i} > R_{\text{sum},i-1}$. Otherwise, the successive allocation of the data streams is stopped and the allocation of the $i^\text{th}$ data stream is undone. The total number of data streams $d$ that can be allocated is limited by the number of RF chains $N_\text{RF}$ at the BS, which is equal to the number of antenna elements in case of fully digital precoding.}

LISA for the traditional fully digital precoding is summarized in Algorithm~\ref{algo:LISA}. The choice of the effective precoder $\boldsymbol{P}_{\text{eff},i}$ containing the precoders $\boldsymbol{p}_{j}$ of the first $i$ allocated data streams $j=1,2,\ldots,i$ according to (\ref{eq:P_eff_i}), ensures that, in the end, after allocating the last data stream $d$, the product of the composite channel matrix $\boldsymbol{H}_{\text{comp},d}$ containing the effective channels $\boldsymbol{g}_i^{\h}\boldsymbol{H}_{\pi\left(i\right)}$ of all $d$ data streams $i=1,2,\ldots,d$ and the effective precoder 
\begin{equation}
\boldsymbol{P}_{\text{eff},d}
=\boldsymbol{Q}_d\boldsymbol{L}_d^{-1}\boldsymbol{\Lambda}_d\boldsymbol{\Gamma}_d^{\frac{1}{2}}
\label{eq:P_eff_d}
\end{equation}
containing the precoders $\boldsymbol{p}_{j}$ of all $d$ data streams $j=1,2,\ldots,d$ is the diagonal matrix $ \boldsymbol{H}_{\text{comp},d}\boldsymbol{P}_{\text{eff},d}=\boldsymbol{\Lambda}_d\boldsymbol{\Gamma}_d^{\frac{1}{2}} $
such that its off-diagonal elements $\boldsymbol{g}_i^{\h}\boldsymbol{H}_{\pi\left(i\right)}\boldsymbol{p}_j$, $j \neq i$, are $0$ and, for each data stream $i$, the interference from all other data streams $j \neq i$ is suppressed.
Finally, the channel is diagonalized and decomposed into $d$ scalar interference-free subchannels, 
whose channel gains are the diagonal elements of $\boldsymbol{\Lambda}_d$ whereas the diagonal elements of $\boldsymbol{\Gamma}_d$ are the powers given to them.

\subsubsection*{Remark}
\revised{The two-stage LISA algorithm is clearly establishing a standard zero-forcing precoder solution if the number of transmitter antenna elements is equal to or even larger than the total number of receiver antenna elements of MSs, since in such cases the zero-forcing precoder is unique up to a precoder component in the nullspace of the composite channel of all MSs. However, in the opposite case of less degrees of freedom at the transmitter, which is clearly met by the multiuser scenario in mmWave communications due to the rather limited number of RF chains at the transmitter side, i.e., $N_\text{RF} < K N_\text{MS} $, a user allocation step is required prior to the deployment of any zero-forcing structure. To this end, a subset of at most $N_\text{RF}$ MSs must been selected to which the limited number of data streams shall be assigned, which essentially resembles a combinatorial problem. Multiple assignments of data streams to the same MS can be desirable. The proposed LISA method solves both problems simultaneously. Although the inherent greedy search technique is of a heuristical nature, its excellent performance has been substantiated in previous publications, e.g., by relating the LISA solution to nontrivial lower bounds of the maximally achievable sum rate, cf. \cite{Guthy2009}, and by the derivation of generalized results based on a large system analysis, cf. \cite{GuUtHo10b,GuUtHo13}.}

\revised{A more intuitive understanding of the benefit of the two-stage nature of LISA can be gained by assuming the second stage of LISA replaced by a dirty-paper coding (DPC) technique, cf. \cite{TeUtBaNo05,Tejera2006}. In this case, the effective precoders are aligned with the successively derived right singular vectors of the first stage. Applying them to the composite channel of the selected MSs results in a lower-triangular structure, which is a prerequisite for any successive interference cancellation method and is also known to be beneficial for DPC. Since LISA cannot rely on DPC, the residual interference between the established data streams must be cancelled by an additional zero-forcing step, this is the second stage of LISA. However, the thoroughly constructed lower-triangular structure in the first stage obviously reduces the residual interference such that LISA still shows an excellent performance after the second stage.}


\begin{algorithm}[tb!]
\caption{Linear Successive Allocation (LISA)}
\label{algo:LISA}
\begin{algorithmic}[1]
    \REQUIRE $\left\{\boldsymbol{H}_k\right\}_{k=1}^{K}$, $N_\text{RF}$, $P$ \\
		\STATE\textbf{Initialize:} $d=N_\text{RF}$, $R_{\text{sum},0}=0$, $\boldsymbol{P}_{k}=\left[\thinspace\right]\enspace\forall k$, $\boldsymbol{G}_{k}=\left[\thinspace\right]\enspace\forall k$, 
			$\boldsymbol{T}_1=\mathbf{I}_{N_\text{BS}}$
		\FOR{$i=1,2,\ldots,d$}
				\STATE $\displaystyle\left\{\pi\left(i\right),\boldsymbol{g}_i,\boldsymbol{q}_i\right\}=\argmax_{k\in\left\{1,2,\ldots,K\right\},\boldsymbol{g}\in\mathbb{C}^{N_\text{MS}},\boldsymbol{q}\in\mathbb{C}^{N_\text{BS}}} \left|\boldsymbol{g}^{\h}\boldsymbol{H}_k\boldsymbol{T}_i\boldsymbol{q}\right| \quad \text{s.t.} \quad \left\|\boldsymbol{g}\right\|_2=\left\|\boldsymbol{q}\right\|_2=1$ \label{algo:LISA:line:maximization}
				
				\STATE $\boldsymbol{H}_{\text{comp},i}=\begin{bmatrix}\boldsymbol{g}_1^{\h}\boldsymbol{H}_{\pi\left(1\right)} \\ \boldsymbol{g}_2^{\h}\boldsymbol{H}_{\pi\left(2\right)} \\ \vdots \\ \boldsymbol{g}_i^{\h}\boldsymbol{H}_{\pi\left(i\right)}\end{bmatrix}$, $\boldsymbol{Q}_{i}=\begin{bmatrix}\boldsymbol{q}_1 & \boldsymbol{q}_2 & \ldots & \boldsymbol{q}_i\end{bmatrix}$, $\boldsymbol{L}_i=\boldsymbol{H}_{\text{comp},i}\boldsymbol{Q}_i$
				
				\STATE $\boldsymbol{\Lambda}_i
				=\Diag\left(\lambda_{i,1},\lambda_{i,2},\ldots,\lambda_{i,i}\right)
				=\left[\Diag\left(\left(\boldsymbol{L}_i^{-1}\right)^{\h}\boldsymbol{L}_i^{-1}\right)\right]^{-\frac{1}{2}}$
				
				\STATE $\displaystyle\left\{\gamma_{i,j}\right\}_{j=1}^{i}=\argmax_{\left\{\gamma_{j}\right\}_{j=1}^{i}} \sum_{j=1}^{i}\log_2\left(1+\gamma_j\lambda_{i,j}^2\right) \quad \text{s.t.} \quad \sum_{j=1}^{i}\gamma_{j} \leq P,\quad \gamma_{j}\geq0 \enspace \forall j$
				\STATE $\displaystyle R_{\text{sum},i}=\sum_{j=1}^{i}\log_2\left(1+\gamma_{i,j}\lambda_{i,j}^2\right)$
				\IF{$R_{\text{sum},i} > R_{\text{sum},i-1}$}
				\STATE $\boldsymbol{\Gamma}_i=\Diag\left(\gamma_{i,1},\gamma_{i,2},\ldots,\gamma_{i,i}\right)$, $\boldsymbol{P}_{\text{eff},i}=\begin{bmatrix}\boldsymbol{p}_1 & \boldsymbol{p}_2 & \ldots & \boldsymbol{p}_i\end{bmatrix}=\boldsymbol{Q}_i\boldsymbol{L}_i^{-1}\boldsymbol{\Lambda}_i\boldsymbol{\Gamma}_i^{\frac{1}{2}}$				
				\STATE $\boldsymbol{T}_{i+1}=\boldsymbol{T}_i-\boldsymbol{q}_i\boldsymbol{q}_i^{\h}$
				\ELSE
				\STATE $d=i-1$
				\STATE \textbf{break}
				\ENDIF
    \ENDFOR \label{algo:LISA:line:begininsert}
		\FOR{$i=1,2,\ldots,d$} \label{algo:LISA:line:endinsert}
				\STATE $\boldsymbol{P}_{\pi\left(i\right)}=\begin{bmatrix}\boldsymbol{P}_{\pi\left(i\right)}&\boldsymbol{p}_i\end{bmatrix}$, $\boldsymbol{G}_{\pi\left(i\right)}=\begin{bmatrix}\boldsymbol{G}_{\pi\left(i\right)}&\boldsymbol{g}_i\end{bmatrix}$

		\ENDFOR
    \ENSURE $\left\{\boldsymbol{P}_k\right\}_{k=1}^{K}$, $\left\{\boldsymbol{G}_k\right\}_{k=1}^{K}$
\end{algorithmic}
\end{algorithm}

\subsection{LISA for Hybrid Precoding (H-LISA)}

For hybrid precoding, the precoders $\boldsymbol{P}_k
$ for the MSs $k=1,2,\ldots,K$ have to have the special structure in (\ref{eq:P_k}). Since the precoder $\boldsymbol{p}_i$ of the $i^\text{th}$ data stream is an element of the precoder $\boldsymbol{P}_{\pi\left(i\right)}$ for the MS $\pi\left(i\right)$, to which the $i^\text{th}$ data stream is allocated, $\boldsymbol{p}_i$ inherits the special structure $\boldsymbol{p}_i=\boldsymbol{P}_{\text{A}}\boldsymbol{p}_{\text{D},i}$ from $\boldsymbol{P}_{\pi\left(i\right)}=\boldsymbol{P}_{\text{A}}\boldsymbol{P}_{\text{D},\pi\left(i\right)}$, where $\boldsymbol{p}_{\text{D},i}$ is an element of $\boldsymbol{P}_{\text{D},\pi\left(i\right)}$. Consequently, the effective precoder $\boldsymbol{P}_{\text{eff},d}$, whose columns are the precoders $\boldsymbol{p}_i$ of all $d$ data streams, has to be of the factored form
\begin{equation}
\boldsymbol{P}_{\text{eff},d}
=\boldsymbol{P}_{\text{A}}\boldsymbol{P}_{\text{D}},
\label{eq:P_eff_d_hybrid}
\end{equation}
where $\boldsymbol{P}_{\text{A}}$ is the analog precoder and $ \boldsymbol{P}_{\text{D}}=\left[\boldsymbol{p}_{\text{D},1},\boldsymbol{p}_{\text{D},2},\ldots,\boldsymbol{p}_{\text{D},d}\right]\in\mathbb{C}^{N_\text{RF} \times d} $
the digital precoder, whose columns are the digital precoders $\boldsymbol{p}_{\text{D},i}$ of the individual data streams. 
The effective precoder $\boldsymbol{P}_{\text{eff},d}$ determined by LISA naturally has the factored form given in (\ref{eq:P_eff_d}), which stems from the two stages taken by LISA for suppressing the interstream interference and perfectly \revised{matches} the hybrid analog and digital architecture. 
In the first step, the composite channel matrix $\boldsymbol{H}_{\text{comp},d}$ is reduced to the lower triangular matrix $\boldsymbol{H}_{\text{comp},d}\boldsymbol{Q}_d=\boldsymbol{L}_d$ by multiplying it with the matrix $\boldsymbol{Q}_d$, which, for each data stream, suppresses the interference from the successively allocated data streams. In the second step, the resulting lower triangular matrix $\boldsymbol{L}_d$ from the first step is transformed into the diagonal matrix $\boldsymbol{L}_d\boldsymbol{L}_d^{-1}\boldsymbol{\Lambda}_d\boldsymbol{\Gamma}_d^{\frac{1}{2}}=\boldsymbol{\Lambda}_d\boldsymbol{\Gamma}_d^{\frac{1}{2}}$ by multiplying it with its inverse $\boldsymbol{L}_d^{-1}$, the diagonal normalization matrix $\boldsymbol{\Lambda}_d$ and the principal square root of the diagonal power loading matrix $\boldsymbol{\Gamma}_d$, which, for each data stream, suppresses the remaining interference from the previously allocated data streams and thus the interference from all other data streams.

Although the effective precoder $\boldsymbol{P}_{\text{eff},d}$ obtained by LISA naturally has the factored form in (\ref{eq:P_eff_d}), it cannot be implemented directly by the hybrid architecture, since the matrix $\boldsymbol{Q}_d$ corresponding to the analog precoder $\boldsymbol{P}_{\text{A}}$ of the hybrid precoder given in (\ref{eq:P_eff_d_hybrid}) does generally  not have constant-modulus entries and cannot be implemented entirely by phase shifters in the analog domain. In order to ensure that the effective precoder $\boldsymbol{P}_{\text{eff},d}$ found by LISA can be implemented by the hybrid architecture, the matrix $\boldsymbol{Q}_d$ is approximated by the analog precoder $\boldsymbol{P}_{\text{A}}$, which keeps only the phases of its entries. More specifically, the element $\left[\boldsymbol{P}_{\text{A}}\right]_{m,n}$ of the matrix $\boldsymbol{P}_{\text{A}}$ in the $m^\text{th}$ row and $n^\text{th}$ column is obtained from the element $\left[\boldsymbol{Q}_d\right]_{m,n}$ of the matrix $\boldsymbol{Q}_d$ in the $m^\text{th}$ row and $n^\text{th}$ column as
\begin{equation}
\left[\boldsymbol{P}_{\text{A}}\right]_{m,n}=\frac{1}{\sqrt{N_\text{BS}}}\exp\left(j\arg\left(\left[\boldsymbol{Q}_d\right]_{m,n}\right)\right).
\label{eq:approximation}
\end{equation}
\revised{In other words, the derived precoding matrix  $\boldsymbol{Q}_d$ is projected onto the feasible set of possible implementations of the analog precoder which is strictly constrained to a network of phase shifters.} \revised{The prefactor $\frac{1}{\sqrt{N_\text{BS}}}$ accounts for a normalization of the respective analog precoding unit. In a realistic implementation case, the prefactor will depend on the choice of the respective power amplifier that drives the analog network of the precoding architecture.}

\revised{In the first step of H-LISA, for each data stream, it is tried to suppress the interference from the successively allocated data streams by multiplying the composite channel matrix $\boldsymbol{H}_{\text{comp},d}$ with the analog precoder $\boldsymbol{P}_{\text{A}}$ instead of $\boldsymbol{Q}_d$ to obtain $\boldsymbol{H}_{\text{comp},d}\boldsymbol{P}_{\text{A}}=\boldsymbol{L}_d\boldsymbol{Q}_d^{\h}\boldsymbol{P}_{\text{A}}$. However, due to the imperfection of $ \boldsymbol{P}_{\text{A}} $, the lower-triangular structure will be slightly destroyed. Consequently, the second step of H-LISA considers the distortion of the lower-triangular structure and the suppression of the interstream interference accordingly, i.e., the resulting matrix $\boldsymbol{L}_d\boldsymbol{Q}_d^{\h}\boldsymbol{P}_{\text{A}}$  is zero forced
by applying $ \left(\boldsymbol{L}_d\boldsymbol{Q}_d^{\h}\boldsymbol{P}_{\text{A}}\right)^{-1}$, which forms the digital part of the hybrid precoder, followed by the diagonal normalization matrix $\boldsymbol{\Lambda}_d$ and the principal square root of a diagonal power loading matrix $\boldsymbol{\Gamma}_d$. The two stages for suppressing the whole interstream interference eventually results in the effective precoder}
\begin{equation}
\boldsymbol{P}_{\text{eff},d}
=\boldsymbol{P}_{\text{A}}\left(\boldsymbol{L}_d\boldsymbol{Q}_d^{\h}\boldsymbol{P}_{\text{A}}\right)^{-1}\boldsymbol{\Lambda}_d\boldsymbol{\Gamma}_d^{\frac{1}{2}},
\label{eq:P_eff_d_hybrid_2}
\end{equation}
where the diagonal matrix $ \boldsymbol{\Lambda}_d =\left[\Diag\left(\left(\boldsymbol{P}_{\text{A}}\left(\boldsymbol{L}_d\boldsymbol{Q}_d^{\h}\boldsymbol{P}_{\text{A}}\right)^{-1}\right)^{\h}\boldsymbol{P}_{\text{A}}\left(\boldsymbol{L}_d\boldsymbol{Q}_d^{\h}\boldsymbol{P}_{\text{A}}\right)^{-1}\right)\right]^{-\frac{1}{2}} $
normalizes the columns of $\boldsymbol{P}_{\text{A}}\left(\boldsymbol{L}_d\boldsymbol{Q}_d^{\h}\boldsymbol{P}_{\text{A}}\right)^{-1}$ to unit norm.
This again ensures that $\boldsymbol{H}_{\text{comp},d}\boldsymbol{P}_{\text{eff},d}=\boldsymbol{\Lambda}_d\boldsymbol{\Gamma}_d^{\frac{1}{2}}$, such that the channel is diagonalized and decomposed into $d$ scalar interference-free subchannels, whose channel gains are the diagonal elements of $\boldsymbol{\Lambda}_d$. The diagonal elements of $\boldsymbol{\Gamma}_d$ are the powers allocated to those scalar subchannels and, given the subchannel gains, can be determined by waterfilling to maximize the sum rate while fulfilling the power constraint.
Now, the effective precoder $\boldsymbol{P}_{\text{eff},d}$ in (\ref{eq:P_eff_d_hybrid_2}) has exactly the same structure as the hybrid precoder in (\ref{eq:P_eff_d_hybrid}), where the digital precoder is given by $ \boldsymbol{P}_{\text{D}} =\left(\boldsymbol{L}_d\boldsymbol{Q}_d^{\h}\boldsymbol{P}_{\text{A}}\right)^{-1}\boldsymbol{\Lambda}_d\boldsymbol{\Gamma}_d^{\frac{1}{2}} $,
and can be implemented by the hybrid architecture.
With the simple approximation by the matrix with constant-modulus entries in (\ref{eq:approximation}), the two stages taken by LISA for suppressing the interstream interference have been mapped to the analog and digital domain of hybrid precoding. The resulting new method for designing the hybrid precoders is called H-LISA.

Since the number of RF chains $N_\text{RF}$ is smaller than the number of BS antenna elements $N_\text{BS}$ in case of hybrid precoding, the total number of data streams $d$ that can be allocated to the MSs is limited by the number of RF chains $N_\text{RF}$ rather than the number of BS antenna elements $N_\text{BS}$, i.e., $d \leq N_\text{RF} < N_\text{BS}$. Hence, the successive allocation of the data streams has to be stopped after allocating the data stream $N_\text{RF}$ at the latest. However, it might be stopped already before the maximum number of data streams $N_\text{RF}$ is reached, i.e., $d < N_\text{RF}$, if there is no further increase in the sum rate when allocating a further data stream. In this case, the analog precoder $\boldsymbol{P}_{\text{A}}\in\mathbb{C}^{N_\text{BS} \times d}$ and the digital precoder $\boldsymbol{P}_{\text{D}}\in\mathbb{C}^{d \times d}$ obtained by H-LISA have $d < N_\text{RF}$ columns and rows, respectively. 
This means that, effectively, only $d$ out of the $N_\text{RF}$ available RF chains are used since each column of $\boldsymbol{P}_{\text{A}}$ and each row of $\boldsymbol{P}_{\text{D}}$ is dedicated to one RF chain.

In order to extend LISA described by Algorithm~\ref{algo:LISA} to H-LISA, 
the lines from Algorithm~\ref{algo:LISA_to_LISA_Hybrid} have to be inserted between the lines~$\ref{algo:LISA:line:begininsert}$ and $\ref{algo:LISA:line:endinsert}$.
\begin{algorithm}
\caption{Extension of LISA to H-LISA}
\label{algo:LISA_to_LISA_Hybrid}
\begin{algorithmic}
\FOR{$m=1,2,\ldots,N_\text{BS}$}
		\FOR{$n=1,2,\ldots,d$}
				\STATE $\left[\boldsymbol{P}_{\text{A}}\right]_{m,n}
=\frac{1}{\sqrt{N_\text{BS}}}\exp\left(j\arg\left(\left[\boldsymbol{Q}_d\right]_{m,n}\right)\right)$
			\ENDFOR
		\ENDFOR
		
		\STATE $\boldsymbol{\Lambda}_d
		=\Diag\left(\lambda_{d,1},\lambda_{d,2},\ldots,\lambda_{d,d}\right)=\left[\Diag\left(\left(\boldsymbol{P}_{\text{A}}\left(\boldsymbol{L}_d\boldsymbol{Q}_d^{\h}\boldsymbol{P}_{\text{A}}\right)^{-1}\right)^{\h}\boldsymbol{P}_{\text{A}}\left(\boldsymbol{L}_d\boldsymbol{Q}_d^{\h}\boldsymbol{P}_{\text{A}}\right)^{-1}\right)\right]^{-\frac{1}{2}}$
		\STATE $\displaystyle\left\{\gamma_{d,j}\right\}_{j=1}^{d}=\argmax_{\left\{\gamma_{j}\right\}_{j=1}^{d}} \sum_{j=1}^{d}\log_2\left(1+\gamma_j\lambda_{d,j}^2\right) \quad \text{s.t.} \quad \sum_{j=1}^{d}\gamma_{j} \leq P,\quad \gamma_{j}\geq0 \enspace \forall j$
				\STATE $\boldsymbol{\Gamma}_d=\Diag\left(\gamma_{d,1},\gamma_{d,2},\ldots,\gamma_{d,d}\right)$, $\boldsymbol{P}_{\text{eff},d}=\begin{bmatrix}\boldsymbol{p}_1 & \boldsymbol{p}_2 & \ldots & \boldsymbol{p}_d\end{bmatrix}=\boldsymbol{P}_{\text{A}}\left(\boldsymbol{L}_d\boldsymbol{Q}_d^{\h}\boldsymbol{P}_{\text{A}}\right)^{-1}\boldsymbol{\Lambda}_d\boldsymbol{\Gamma}_d^{\frac{1}{2}}$				
\end{algorithmic}
\end{algorithm}

\section{Low-Complexity Version of LISA and H-LISA}\label{sec4}
For each allocation of a data stream, and the determination of the corresponding equalizer and precoder according to (\ref{eq:LISA_allocation}), the SVDs or at least the maximum singular values and the corresponding singular vectors of the $K$ projected channel matrices $\boldsymbol{H}_k\boldsymbol{T}_i$ have to be computed, which still results in large computational complexity. Therefore, we propose a low-complexity version of LISA, which circumvents the computation of the SVDs 
by exploiting the special structure of the channel matrices $\boldsymbol{H}_k$ in (\ref{eq:H_k}) according to the geometric channel model used for the mmWave channels. 

\revised{Since the equalizers of the data streams are corresponding to the left singular vectors of the projected channel matrices, cf. (\ref{eq:LISA_allocation}), all successively assigned equalizers of data streams allocated to the same MS are inherently orthogonal, i.e., the equalizer $\boldsymbol{g}_i$ of the $i^\text{th}$ data stream allocated to the MS $\pi\left(i\right)$ lies in the nullspace of the row vectors $\boldsymbol{g}_j^{\h}$ of the data streams $j<i$ previously allocated to the same MS $\pi\left(j\right)=\pi\left(i\right)$, i.e., $\boldsymbol{g}_i\in\operatorname{null}\left\{\boldsymbol{g}_j^{\h}:j=1,2,\ldots,i-1 \wedge \pi\left(j\right)=\pi\left(i\right)\right\}$.} 
\revised{In order to maintain this property also in more general cases, e.g., the following derivation of a low-complex version of LISA and H-LISA, the optimization problem is rewritten by adding an additional constraint. To this end, 
the projector $\boldsymbol{S}_{k,i}\in\mathbb{C}^{N_\text{MS} \times N_\text{MS}}$ onto the nullspace $\operatorname{null}\left\{\boldsymbol{g}_j^{\h}:j=1,2,\ldots,i-1 \wedge \pi\left(j\right)=k\right\}$ is introduced,} such that (\ref{eq:LISA_allocation}) can be equivalently stated as
\begin{equation}
\left\{\pi\left(i\right),\boldsymbol{g}_i,\boldsymbol{q}_i\right\}=
\argmax_{\substack{k\in\left\{1,2,\ldots,K\right\}\\\boldsymbol{g}\in\mathbb{C}^{N_\text{MS}},\boldsymbol{q}\in\mathbb{C}^{N_\text{BS}}}}
\left|\boldsymbol{g}^{\h}\boldsymbol{S}_{k,i}\boldsymbol{H}_k\boldsymbol{T}_i\boldsymbol{q}\right| \quad \text{s.t.} \quad 
\left\|\boldsymbol{g}\right\|_2=\left\|\boldsymbol{q}\right\|_2=1.
\label{eq:LISA_allocation_3}
\end{equation}
Similarly to $\boldsymbol{T}_i$, the projectors $\boldsymbol{S}_{k,i}$ for the different MSs $k=1,2,\ldots,K$ can be computed recursively, starting at $\boldsymbol{S}_{k,1}=\mathbf{I}_{N_\text{MS}}$:
\begin{equation}
\boldsymbol{S}_{k,i+1}=
\begin{cases}
\boldsymbol{S}_{k,i}-\boldsymbol{g}_{i}\boldsymbol{g}_{i}^{\h}, & k=\pi\left(i\right) \\
\boldsymbol{S}_{k,i}, & k\neq\pi\left(i\right).
\end{cases}
\end{equation}
Now, applying the special structure of the channel matrix $\boldsymbol{H}_k$ according to the geometric channel model for mmWave channels in (\ref{eq:H_k})%
, the projected channel matrix $\boldsymbol{S}_{k,i}\boldsymbol{H}_k\boldsymbol{T}_i$ in the objective 
of the optimization problem (\ref{eq:LISA_allocation_3}) 
results in $ \boldsymbol{S}_{k,i}\boldsymbol{H}_k\boldsymbol{T}_i=\boldsymbol{S}_{k,i}\sqrt{\frac{N_\text{BS}N_\text{MS}}{L_k}}\sum_{\ell=1}^{L_k}\alpha_{k,\ell}\boldsymbol{a}_\text{MS}\left(\phi^\text{MS}_{k,\ell},\theta^\text{MS}_{k,\ell}\right) $ $ \times \boldsymbol{a}_\text{BS}^{\h}\left(\phi^\text{BS}_{k,\ell},\theta^\text{BS}_{k,\ell}\right)\boldsymbol{T}_i $. 
It can be reformulated as
\begin{equation}
\begin{split}
\boldsymbol{S}_{k,i}\boldsymbol{H}_k\boldsymbol{T}_i 
&=\sum_{\ell=1}^{L_k}\alpha_{k,\ell,i}\boldsymbol{a}_{\text{MS},k,\ell,i}\boldsymbol{a}_{\text{BS},k,\ell,i}^{\h},
\end{split}
\label{eq:LISA_projected_channel_matrix_2}
\end{equation}
where
\begin{equation}
\boldsymbol{a}_{\text{BS},k,\ell,i}
=\frac{\boldsymbol{T}_i\boldsymbol{a}_\text{BS}\left(\phi^\text{BS}_{k,\ell},\theta^\text{BS}_{k,\ell}\right)}{\left\|\boldsymbol{T}_i\boldsymbol{a}_\text{BS}\left(\phi^\text{BS}_{k,\ell},\theta^\text{BS}_{k,\ell}\right)\right\|_2}
\end{equation}
is the normalized array response vector $\boldsymbol{a}_\text{BS}\left(\phi^\text{BS}_{k,\ell},\theta^\text{BS}_{k,\ell}\right)$ at the BS for the $\ell^\text{th}$ path to the $k^\text{th}$ MS projected onto the nullspace of the effective channels $\boldsymbol{g}_j^{\h}\boldsymbol{H}_{\pi\left(j\right)}$ of the previously allocated data streams $j<i$, 
and likewise
\begin{equation}
\boldsymbol{a}_{\text{MS},k,\ell,i}
=\frac{\boldsymbol{S}_{k,i}\boldsymbol{a}_\text{MS}\left(\phi^\text{MS}_{k,\ell},\theta^\text{MS}_{k,\ell}\right)}{\left\|\boldsymbol{S}_{k,i}\boldsymbol{a}_\text{MS}\left(\phi^\text{MS}_{k,\ell},\theta^\text{MS}_{k,\ell}\right)\right\|_2}
\end{equation}
is the normalized array response vector $\boldsymbol{a}_\text{MS}\left(\phi^\text{MS}_{k,\ell},\theta^\text{MS}_{k,\ell}\right)$ at the $k^\text{th}$ MS for the $\ell^\text{th}$ path from the BS projected onto the nullspace of the row vectors $\boldsymbol{g}_j^{\h}$ of the data streams $j<i$ previously allocated to the same MS $k$. The additional factor
\begin{align*} \alpha_{k,\ell,i}&=\sqrt{\frac{N_\text{BS}N_\text{MS}}{L_k}} \alpha_{k,\ell} \left\|\boldsymbol{S}_{k,i}\boldsymbol{a}_\text{MS}\left(\phi^\text{MS}_{k,\ell},\theta^\text{MS}_{k,\ell}\right)\right\|_2 \left\|\boldsymbol{T}_i\boldsymbol{a}_\text{BS}\left(\phi^\text{BS}_{k,\ell},\theta^\text{BS}_{k,\ell}\right)\right\|_2
\end{align*}
weights the outer product of the projected array response vectors $\boldsymbol{a}_{\text{MS},k,\ell,i}$ and $\boldsymbol{a}_{\text{BS},k,\ell,i}$ for the MS $k$ and the path $\ell$.

In order to avoid solving the optimization problem (\ref{eq:LISA_allocation_3}) by computing the maximum singular values and the corresponding left and right singular vectors of the $K$ projected channel matrices $\boldsymbol{S}_{k,i}\boldsymbol{H}_k\boldsymbol{T}_i$, their special structure is exploited. 

{\color{black}{To this end, the optimization problem (\ref{eq:LISA_allocation_3}) is replaced by
\begin{align}
\left\{\pi\left(i\right),\ell\left(i\right)\right\}
&=\argmax_{k\in\left\{1,2,\ldots,K\right\},\ell\in\left\{1,2,\ldots,L_k\right\}}
\left|\alpha_{k,\ell,i}\right|,
\label{eq:low-complexity_user_selection} \\
\boldsymbol{g}_i & =\boldsymbol{a}_{\text{MS},\pi\left(i\right),\ell\left(i\right),i}, \nonumber \\
\boldsymbol{q}_i & =\frac{\boldsymbol{T}_i\boldsymbol{H}^{\h}_{\pi\left(i\right)}\boldsymbol{g}_i}
{\left\|\boldsymbol{T}_i\boldsymbol{H}^{\h}_{\pi\left(i\right)}\boldsymbol{g}_i\right\|_2}, \nonumber
\end{align}
i.e., by choosing $\pi\left(i\right)$ and $\ell\left(i\right)$ corresponding to the $k^\text{th}$ MS and its $\ell^\text{th}$ path with the largest weight factor $\alpha_{k,\ell,i}$, by defining the equalizer $\boldsymbol{g}_i$ as the projected array response vector $\boldsymbol{a}_{\text{MS},\pi\left(i\right),\ell\left(i\right),i}$ and matching the precoder $\boldsymbol{q}_i$ to the equalizer $\boldsymbol{g}_i$ as in the original version of LISA. This ensures that the
precoder $\boldsymbol{q}_i$ of the $i^\text{th}$ data stream again lies in the nullspace of the effective channels $\boldsymbol{g}_j^{\h}\boldsymbol{H}_{\pi\left(j\right)}$ of the previously allocated data streams $j<i$, $\operatorname{null}\left\{\boldsymbol{g}_j^{\h}\boldsymbol{H}_{\pi\left(j\right)}\right\}_{j=1}^{i-1}$, and the product of the composite channel matrix $\boldsymbol{H}_{\text{comp},i}$ and the matrix $\boldsymbol{Q}_{i}$, whose orthonormal columns are the precoders $\boldsymbol{q}_j$ of the first $i$ data streams $j=1,2,\ldots,i$, is again a lower triangular matrix $\boldsymbol{L}_{i}$, such that for each data stream the interference from the successively allocated data streams is suppressed.
}}

{\begin{algorithm}
\caption{Low-Complexity LISA}
\label{algo:Low-Complexity_LISA}
\begin{algorithmic}[1]
    \REQUIRE
			$N_\text{BS}$, $N_\text{MS}$, $N_\text{RF}$, $P$, $\left\{\alpha_{k,\ell}\right\}$, $\left\{\phi^\text{BS}_{k,\ell}\right\}$, $\left\{\theta^\text{BS}_{k,\ell}\right\}$,
			$\left\{\phi^\text{MS}_{k,\ell}\right\}$, $\left\{\theta^\text{MS}_{k,\ell}\right\}$, $k=1,\ldots,K$, $\ell=1,\ldots,L_k$
				
		\STATE\textbf{Initialize:} $d=N_\text{RF}$, $R_{\text{sum},0}=0$, $\boldsymbol{P}_{k}=\left[\thinspace\right]\enspace\forall k$, $\boldsymbol{G}_{k}=\left[\thinspace\right]\enspace\forall k$, $\boldsymbol{T}_1=\mathbf{I}_{N_\text{BS}}$, $\boldsymbol{S}_{k,1}=\mathbf{I}_{N_\text{MS}}\enspace\forall k$
		\FOR{$k=1,2,\ldots,K$}
				\STATE $\boldsymbol{H}_k=\displaystyle\sqrt{\frac{N_\text{BS}N_\text{MS}}{L_k}}\sum_{\ell=1}^{L_k}\alpha_{k,\ell}\boldsymbol{a}_\text{MS}\left(\phi^\text{MS}_{k,\ell},\theta^\text{MS}_{k,\ell}\right)\boldsymbol{a}_\text{BS}^{\h}\left(\phi^\text{BS}_{k,\ell},\theta^\text{BS}_{k,\ell}\right)$
		\ENDFOR
		\FOR{$i=1,2,\ldots,d$}
				\STATE $\displaystyle\left\{\pi\left(i\right),\ell\left(i\right)\right\}
				=\argmax_{\substack{k\in\left\{1,2,\ldots,K\right\}\\\ell\in\left\{1,2,\ldots,L_k\right\}}}\left|
				\sqrt{\frac{N_\text{BS}N_\text{MS}}{L_k}}\alpha_{k,\ell}
				\left\|\boldsymbol{S}_{k,i}\boldsymbol{a}_\text{MS}\left(\phi^\text{MS}_{k,\ell},\theta^\text{MS}_{k,\ell}\right)\right\|_2
				\left\|\boldsymbol{T}_i\boldsymbol{a}_\text{BS}\left(\phi^\text{BS}_{k,\ell},\theta^\text{BS}_{k,\ell}\right)\right\|_2
				\right|$ \label{algo:Low-Complexity_LISA:line:maximization}
				\STATE $\boldsymbol{g}_i
				=\frac{\boldsymbol{S}_{\pi\left(i\right),i}\boldsymbol{a}_\text{MS}\left(\phi^\text{MS}_{\pi\left(i\right),\ell\left(i\right)},\theta^\text{MS}_{\pi\left(i\right),\ell\left(i\right)}\right)}
				{\left\|\boldsymbol{S}_{\pi\left(i\right),i}\boldsymbol{a}_\text{MS}\left(\phi^\text{MS}_{\pi\left(i\right),\ell\left(i\right)},\theta^\text{MS}_{\pi\left(i\right),\ell\left(i\right)}\right)\right\|_2}$\label{algo:Low-Complexity_LISA:line:g_i},
				$\boldsymbol{q}_i
				=\frac{\boldsymbol{T}_i\boldsymbol{H}^{\h}_{\pi\left(i\right)}\boldsymbol{g}_i}
				{\left\|\boldsymbol{T}_i\boldsymbol{H}^{\h}_{\pi\left(i\right)}\boldsymbol{g}_i\right\|_2}$,  $\boldsymbol{H}_{\text{comp},i}=\begin{bmatrix}\boldsymbol{g}_1^{\h}\boldsymbol{H}_{\pi\left(1\right)} \\ \boldsymbol{g}_2^{\h}\boldsymbol{H}_{\pi\left(2\right)} \\ \vdots \\ \boldsymbol{g}_i^{\h}\boldsymbol{H}_{\pi\left(i\right)}\end{bmatrix}$
				
        \STATE $\boldsymbol{Q}_{i}=\begin{bmatrix}\boldsymbol{q}_1 & \boldsymbol{q}_2 & \ldots & \boldsymbol{q}_i\end{bmatrix}$,
				$\boldsymbol{L}_i=\boldsymbol{H}_{\text{comp},i}\boldsymbol{Q}_i$
		\STATE $\boldsymbol{\Lambda}_i
				=\Diag\left(\lambda_{i,1},\lambda_{i,2},\ldots,\lambda_{i,i}\right)
				=\left[\Diag\left(\left(\boldsymbol{L}_i^{-1}\right)^{\h}\boldsymbol{L}_i^{-1}\right)\right]^{-\frac{1}{2}}$
				
				\STATE $\displaystyle\left\{\gamma_{i,j}\right\}_{j=1}^{i}=\argmax_{\left\{\gamma_{j}\right\}_{j=1}^{i}} \sum_{j=1}^{i}\log_2\left(1+\gamma_j\lambda_{i,j}^2\right)\quad\displaystyle\text{s.t.} \quad \sum_{j=1}^{i}\gamma_{j} \leq P,\quad \gamma_{j}\geq0 \enspace \forall j$
				\STATE $\displaystyle R_{\text{sum},i}=\sum_{j=1}^{i}\log_2\left(1+\gamma_{i,j}\lambda_{i,j}^2\right)$
				\IF{$R_{\text{sum},i} > R_{\text{sum},i-1}$}
				\STATE $\boldsymbol{\Gamma}_i=\Diag\left(\gamma_{i,1},\gamma_{i,2},\ldots,\gamma_{i,i}\right)$, $\boldsymbol{P}_{\text{eff},i}=\begin{bmatrix}\boldsymbol{p}_1 & \boldsymbol{p}_2 & \ldots & \boldsymbol{p}_i\end{bmatrix}=\boldsymbol{Q}_i\boldsymbol{L}_i^{-1}\boldsymbol{\Lambda}_i\boldsymbol{\Gamma}_i^{\frac{1}{2}}$				
				\STATE $\boldsymbol{T}_{i+1}=\boldsymbol{T}_i-\boldsymbol{q}_i\boldsymbol{q}_i^{\h}$
				\FOR{$k=1,2,\ldots,K$}
					\IF{$k=\pi\left(i\right)$}
					\STATE $\boldsymbol{S}_{k,i+1}=\boldsymbol{S}_{k,i}-\boldsymbol{g}_{i}\boldsymbol{g}_{i}^{\h}$
					\ELSE
					\STATE $\boldsymbol{S}_{k,i+1}=\boldsymbol{S}_{k,i}$
					\ENDIF
				\ENDFOR
				\ELSE
				\STATE $d=i-1$
				\STATE \textbf{break}
				\ENDIF
    \ENDFOR \label{algo:Low-Complexity_LISA:line:begininsert}
		\FOR{$i=1,2,\ldots,d$} \label{algo:Low-Complexity_LISA:line:endinsert}
				\STATE $\boldsymbol{P}_{\pi\left(i\right)}=\begin{bmatrix}\boldsymbol{P}_{\pi\left(i\right)}&\boldsymbol{p}_i\end{bmatrix}$,
				$\boldsymbol{G}_{\pi\left(i\right)}=\begin{bmatrix}\boldsymbol{G}_{\pi\left(i\right)}&\boldsymbol{g}_i\end{bmatrix}$

		\ENDFOR
    \ENSURE $\left\{\boldsymbol{P}_k\right\}_{k=1}^{K}$, $\left\{\boldsymbol{G}_k\right\}_{k=1}^{K}$
\end{algorithmic}
\end{algorithm}}

The low-complexity version of LISA is summarized in Algorithm~\ref{algo:Low-Complexity_LISA}. Again, 
the lines from Algorithm~\ref{algo:LISA_to_LISA_Hybrid} can be inserted between the lines~$\ref{algo:Low-Complexity_LISA:line:begininsert}$ and $\ref{algo:Low-Complexity_LISA:line:endinsert}$ in order to obtain a low-complexity version of H-LISA. It should be noted that these low-complexity algorithms use the geometric parameters of the channels, i.e., the complex path gains $\alpha_{k,\ell}$, the azimuth and elevation angles of departure $\phi^\text{BS}_{k,\ell}$ and $\theta^\text{BS}_{k,\ell}$ as well as the azimuth and elevation angles of arrival $\phi^\text{MS}_{k,\ell}$ and $\theta^\text{MS}_{k,\ell}$, rather than the channel matrices $\boldsymbol{H}_k$ as inputs. Therefore, during channel estimation, those geometric parameters of the channels are to be estimated.

\begin{table}[!h]
\begin{center}
\renewcommand{\arraystretch}{1.3}
\textcolor{black}{
\caption{{Computational complexity}}
\label{tab:complexity}
\centering
\begin{tabular}{c||c|c|c}
\hline
operations & H-LISA & Low-Complexity H-LISA & 2SMUHPA \\\hline\hline
composite channel matrices & \underline{$\mathcal{O}\left(KdN_\text{BS}N_\text{MS}^2\right)$} & \underline{$\mathcal{O}\left(KLdN_\text{BS}\right)$} & \underline{$\mathcal{O}\left(KN_\text{BS}\left(K+N_\text{MS}\right)\right) $}\\
projections & $\mathcal{O}\left(KdN_\text{BS}N_\text{MS}\right)$ & $\mathcal{O}\left(KLdN_\text{BS}\right)$ & -- \\
zero-forcing & $\mathcal{O}\left(d^2N_\text{BS}\right)$ & $\mathcal{O}\left(d^2N_\text{BS}\right)$ & $\mathcal{O}\left(K^2N_\text{BS}\right)$\\\hline
total & $\mathcal{O}\left(KdN_\text{BS}N_\text{MS}^2\right)$ & $\mathcal{O}\left(KLdN_\text{BS}\right)$ & $\mathcal{O}\left(KN_\text{BS}\left(K+N_\text{MS}\right)\right)$
\end{tabular}}
\end{center}
\end{table}

{\color{black}{Table \ref{tab:complexity} finally summarizes the order of numerical complexity of the major computational steps of each of the discussed multiuser hybrid precoding methods, i.e., the construction of the composite channel matrix, the performed projections and the zero-forcing.  The computational complexity of LISA and H-LISA is clearly dominated by successively computing the SVDs of the $K$ projected channel matrices $\boldsymbol{H}_k\boldsymbol{T}_i$ in order to construct the composite channel matrix $ \boldsymbol{H}_{\text{comp},i} $. For the low-complexity version of LISA and H-LISA, we assume that the number of paths between the BS and each MS is $L$, which is typically small for mmWave channels in contrast to the number of antenna elements deployed at the BS. The numerical complexity is effectively dominated by the computation of the weights $\alpha_{k,\ell,i}$ for all $k=1,2,\ldots,K$, $\ell=1,2,\ldots,L$ and $i=1,2,\ldots,d$. These weights are important for assigning the $i^\text{th}$ data stream to the MS $k$ and}} {\color{black}{the path $\ell$ according to (\ref{eq:low-complexity_user_selection}). The smallest order of numerical complexity can be stated for the 2SMUHPA, where the construction of the composite channel matrix $\hat{\boldsymbol{H}}$ is the most demanding computational step.}}



\section{Modification of H-LISA for Analog Processing at MSs}\label{sec5}

So far, it has been assumed that the MSs have as many RF chains as antenna elements and the whole signal processing takes place in the digital domain. Following the same reasoning as for the reduction of complexity, power consumption and costs at the BS by reducing the number of RF chains, each MS is now equipped with $N^\text{MS}_\text{RF}<N_\text{MS}$ RF chains, where each RF chain is connected with each antenna element via a phase shifter. The equalizer $\boldsymbol{G}_k\in\mathbb{C}^{N_\text{MS} \times d_k}$ is implemented by the resulting network of phase shifters between the $N_\text{RF}^\text{MS}$ RF chains limiting the number of data streams $d_k$ and the $N_\text{MS}$ antenna elements in the analog domain such that $d_k \leq N_\text{RF}^\text{MS}$ and each element of $\boldsymbol{G}_k$ has to fulfill the constant-modulus constraint $\left|\left[\boldsymbol{G}_{k}\right]_{m,n}\right|=\frac{1}{\sqrt{N_\text{MS}}}$.

The following slight modification of H-LISA and its low-complexity version ensures that those requirements are met. First, all MSs to which $N^\text{MS}_\text{RF}$ data streams have already been allocated are excluded from the allocation of a further data stream. More specifically, the set of all MSs $k$, i.e., $\left\{1,2,\ldots,K\right\}$, from which the MS $\pi\left(i\right)$ for the $i^\text{th}$ data stream is selected in line~$\ref{algo:LISA:line:maximization}$ of Algorithm~\ref{algo:LISA} and line~$\ref{algo:Low-Complexity_LISA:line:maximization}$ of Algorithm~\ref{algo:Low-Complexity_LISA}, is restricted to the set of MSs $k$ whose current number of data streams $d_{k,i-1}$ is less than $N_\text{RF}^\text{MS}$, i.e., $\left\{k\in\left\{1,2,\ldots,K\right\}:d_{k,i-1}<N_\text{RF}^\text{MS}\right\}$, such that $d_k \leq N_\text{RF}^\text{MS}$.
Second, the equalizer $\boldsymbol{g}_i$ for the $i^\text{th}$ data stream, which has been found in line~$\ref{algo:LISA:line:maximization}$ of Algorithm~\ref{algo:LISA} and line~$\ref{algo:Low-Complexity_LISA:line:g_i}$ of Algorithm~\ref{algo:Low-Complexity_LISA}, is set to its simple approximation $\boldsymbol{g}'_i\in\mathbb{C}^{N_\text{MS}}$, whose $m^\text{th}$ element is given by $ \left[\boldsymbol{g}'_{i}\right]_{m}=\frac{1}{\sqrt{N_\text{MS}}}\exp\left(j\arg\left(\left[\boldsymbol{g}_i\right]_{m}\right)\right) $. Similarly to the approximation of $\boldsymbol{Q}_d$ by the analog precoder $\boldsymbol{P}_{\text{A}}$ in (\ref{eq:approximation}), it keeps only the phases of all entries and normalizes them to the same modulus $\frac{1}{\sqrt{N_\text{MS}}}$ such that, in the end, the equalizers $\boldsymbol{G}_k$ formed from those equalizers $\boldsymbol{g}_i$ of the individual data streams fulfill the constant-modulus constraint $\left|\left[\boldsymbol{G}_{k}\right]_{m,n}\right|=\frac{1}{\sqrt{N_\text{MS}}}$. \revised{Although this projection onto the feasible set of possible implementations of the analog equalizer would typically require a hybrid architecture with a subsequent digital part, we refrain from any digital postprocessing by shifting the necessary correction of the non-ideal analog part at the MS to a respective adaptation of the precoders at the transmitter, in other words, the precoders take over the digital postprocessing at the MSs taking into account the approximated equalizers $\boldsymbol{g}_i^\prime$.}

\section{Numerical Results}\label{sec6}

In order to evaluate the performance of the methods for hybrid precoding described in the previous section, we have conducted some simulations, whose numerical results are presented in this section.

For the simulations, a system consisting of $K=8$ MSs equipped with $N_\text{MS}$-element UPAs and a BS, which is equipped with an $8\times8$ UPA ($N_\text{BS}=64$) and $N_\text{RF}=8$ RF chains supporting up to $8$ data streams, is considered.
The channel matrices $\boldsymbol{H}_k$ have the special structure in (\ref{eq:H_k}) according to the adopted geometric channel model. It is assumed that there are $L$ paths between the BS and each MS, i.e., $L_k=L\enspace\forall k$. The azimuth angles of arrival and departure $\phi^\text{MS}_{k,\ell}$ and $\phi^\text{BS}_{k,\ell}$ are selected uniformly at random from the interval $\left[0^\circ,360^\circ\right]$ while the elevation angles of arrival and departure $\theta^\text{MS}_{k,\ell}$ and $\theta^\text{BS}_{k,\ell}$ are selected uniformly at random from the interval $\left[-90^\circ,90^\circ\right]$.
The average sum rate obtained in $1000$ Monte Carlo runs according to (\ref{eq:sum_rate}) serves as performance measure and the SNR is defined as $ \text{SNR} =P $.

In addition, the average sum rate achieved by the hybrid precoding methods is also compared to that achieved by LISA and its low-complexity version for fully digital precoding, and to the capacity of the MIMO broadcast channel between the BS and the MSs. Although LISA and its low-complexity version for fully digital precoding do not have to fulfill the constant-modulus constraints resulting from the analog processing, they are also restricted to allocate at most $8$ data streams as their hybrid counterparts for a fair comparison. The average maximum sum capacity achievable with dirty-paper coding (DPC), which is an upper bound on the average sum rate, is computed with sum power iterative waterfilling proposed in \cite{Jindal2005}.

\begin{figure}[h!]
\centering
\subfloat[$L=1$]{
\input{Sum_Rate_vs_SNR_GeometricUniform_BSUPA_MSULA_L1_K8_N_BS64_N_MS1_2017-05-08-22-08-44.tikz}
}
\qquad
\subfloat[$L=3$]{
\input{Sum_Rate_vs_SNR_GeometricUniform_BSUPA_MSULA_L3_K8_N_BS64_N_MS1_2017-05-08-21-35-08.tikz}
}
\caption{\revised{Average Sum Rate vs. SNR for $N_\text{BS}=64$, $N_\text{MS}=1$, $N_\text{RF}=8$, $K=8$, $L\in\{1,3\}$.}}
\label{fig:Simulation}
\end{figure}
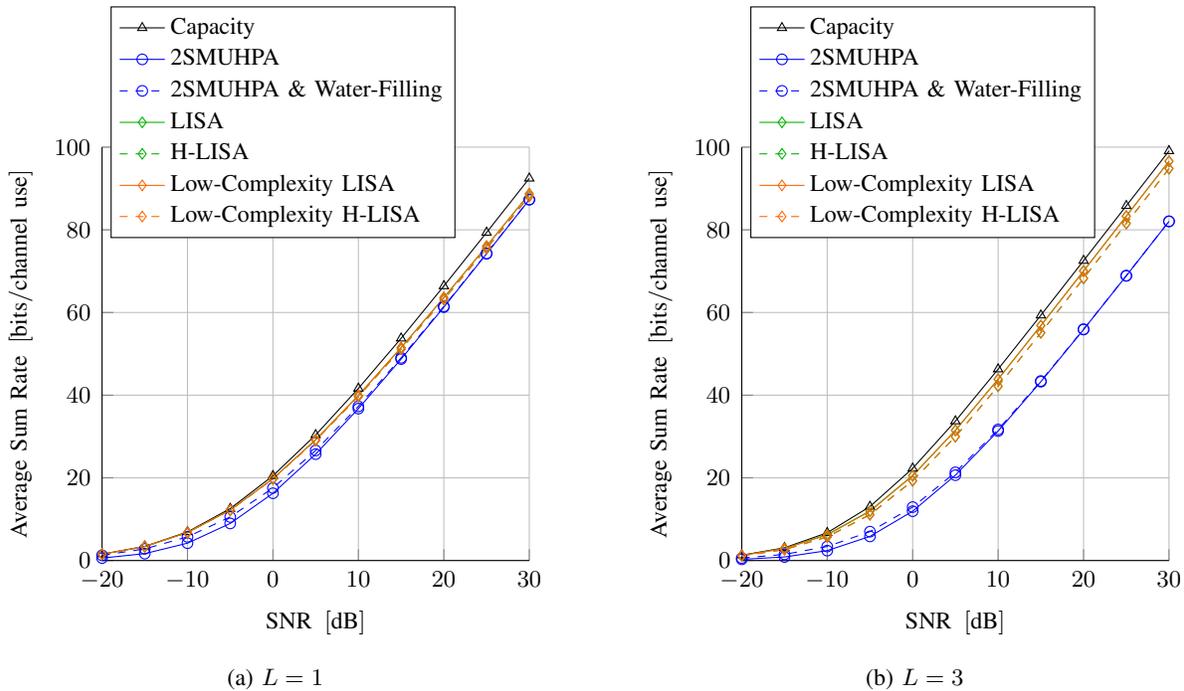

\begin{figure}[h!]\centering
\input{Channel_Gains_vs_SNR_GeometricUniform_BSUPA_MSULA_L3_K8_N_BS64_N_MS1_2016-05-07-11-01-20.tikz}
\vspace{-4mm}
\caption{Channel gains for $N_\text{BS}=64$, $N_\text{MS}=1$, $N_\text{RF}=8$, $K=8$, $L=3$, $\text{SNR}=0\text{ dB}$.}
\label{fig:Simulation_4}
\end{figure}
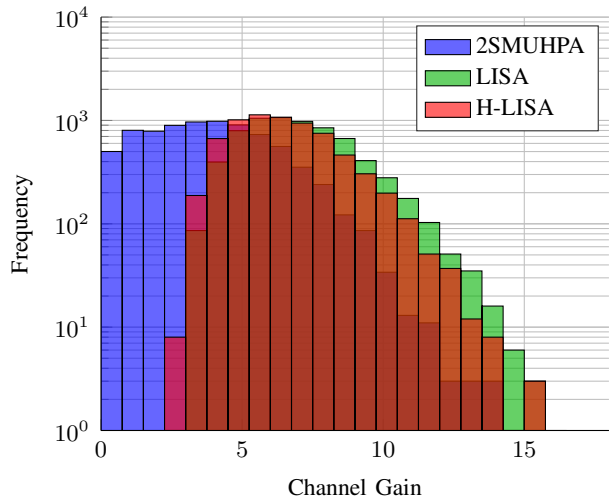

In Fig.~\ref{fig:Simulation}, the average sum rate is plotted over the SNR for single-antenna MSs ($N_\text{MS}=1$) and \revised{single- and multipath channels ($L \in \{1, 3\}$).} The 2SMUHPA with waterfilling performs better than its original version without waterfilling. LISA performs better than the 2SMUHPA with waterfilling and comes close to the capacity. For single-path channels, the simple \revised{heuristic} approximation by the matrix with constant-modulus entries in H-LISA does not lead to a significant performance degradation, and the absolute values of the complex path gains $\alpha_{k,1}$ are the maximum singular values of the channel matrices $\boldsymbol{H}_k$ and the array response vectors are scaled versions of the corresponding singular vectors such that there is no difference between the original and low-complexity versions of LISA and H-LISA, i.e., orange and green curves in \revised{Fig.~\ref{fig:Simulation}a} are matching completely. At an SNR of $0$ dB, the average sum rate is approximately $18$ and $16$ bits per channel use for the 2SMUHPA with and without waterfilling, respectively. LISA, H-LISA and their low-complexity versions, however, achieve an average sum rate of approximately $20$ bits per channel use, which is only slightly smaller than the capacity of approximately $21$ bits per channel use.
\revised{If there are multiple paths between the BS and each MS, the performance gap between the two versions of the 2SMUHPA and the four versions of LISA becomes larger as can be seen in Fig.~\ref{fig:Simulation}b. The reason for the correspondence between increasing gains and the growing number of propagation paths lies in the nature of the 2SMUHPA, which, in contrast to the proposed methods, selects the analog beamforming based on the array response vectors corresponding to the paths between the BS and MS irrespective of any interference that might occur. On the contrary, LISA exploits the increased number of degrees of freedom corresponding to the growing number of propagation paths already within the analog part for maximizing the effective transmission gains subject to the desired interference suppression. This explains the slight degradation of the conventional 2-stage hybrid precoding even in case of a single propagation path.}
The average sum rate of LISA is still close to the capacity. However, the simple approximation by the matrix with constant-modulus entries in H-LISA leads to a larger performance degradation. The low-complexity versions of LISA and H-LISA achieve almost the same average sum rate as LISA and H-LISA, respectively. The 2SMUHPA with and without waterfilling achieves an average sum rate of approximately $13$ and $12$ bits per channel use, respectively, at an SNR of $0$ dB. With approximately $19$ bits per channel use, the average sum rate of the two versions of H-LISA is much larger. The average sum rate of the two versions of LISA is approximately $20.5$ bits per channel use and close to the capacity of approximately $22$ bits per channel use.
Moreover, the distribution of the effective channel gains of H-LISA shows that the \revised{heuristic} approximation by the matrix with constant-modulus entries in the first step of H-LISA largely preserves the large gains of LISA, which makes it a promising method for hybrid precoding. The effective channel gains of the scalar interference-free subchannels \revised{are} illustrated by the histogram in Fig.~\ref{fig:Simulation_4} at an SNR of $0$ dB.
\revised{All methods achieve the maximal slope of the capacity curve at high SNR (degrees of freedom) which is equal to $ \min(N_\text{RF},K\times N_\text{MS},K\times L) = 8 $.}

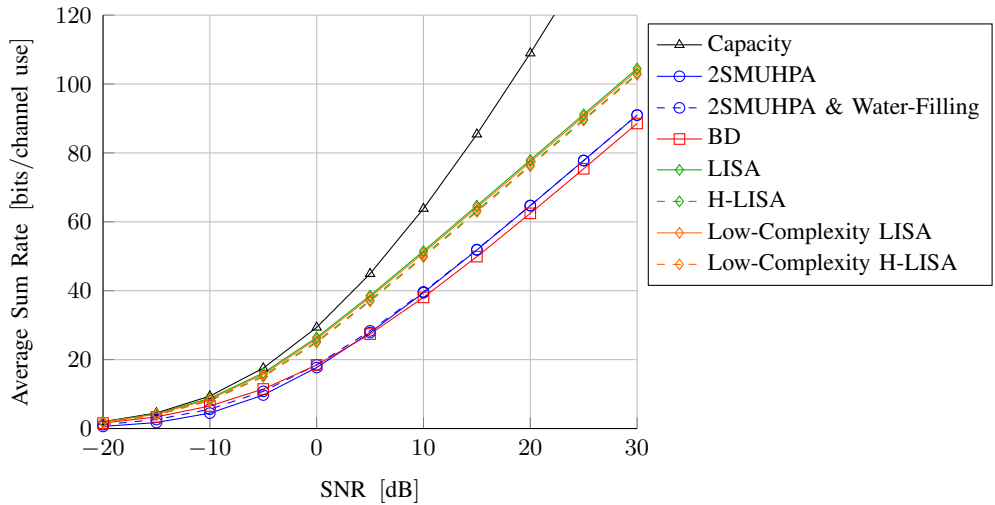
\begin{figure}[h!]\centering
\input{Sum_Rate_vs_SNR_GeometricUniform_BSUPA_MSULA_L3_K8_N_BS64_N_MS2_N_RF_MS2_2017-05-08-22-17-53.tikz}
\\ \vspace{-4mm}
\caption{Average Sum Rate vs. SNR for $N_\text{BS}=64$, $N_\text{MS}=2$, $N_\text{RF}=8$, $K=8$, $L=3$.}
\label{fig:Simulation_6}
\end{figure}

\revised{In Fig.~\ref{fig:Simulation_6}, we now consider MSs with two antenna elements each. 
For this scenario, the average sum rate achievable with Block-Diagonalization (BD) is included. The method has been proposed for the traditional fully digital precoding in\cite{Spencer2004}.} As in \cite{Shen2006}, we assume that each MS receives either no or the maximum number of data streams, i.e., $N_\text{MS}=2$, such that the maximum number of MSs that can be simultaneously supported by BD is 
$\hat{K}=\frac{N_\text{RF}}{N_\text{MS}}=4$. An exhaustive search over all possible sets of $1 \leq i \leq \hat{K}$ MSs is performed, where BD is applied for each of those sets, to select the BD solution with the largest sum rate in the end. The so obtained average sum rate of BD is significantly smaller than that of the LISA-based methods especially at high SNR. At an SNR of $0$ dB, the average sum rate of approximately $26$ bits per channel use achieved by LISA and its low-complexity version, and the average sum rate of approximately $25$ bits per channel use achieved by H-LISA and its low-complexity version are much larger than the average sum rate of approximately $18$ bits per channel use achieved by the 2SMUHPA with or without waterfilling and BD, and smaller than the channel capacity of approximately $29$ bits per channel use at an SNR of $0$ dB. \revised{The increasing performance gap between the 2SMUHPA and LISA is due to the almost optimal greedy selection process of LISA which decides how many data streams are allocated to which MS. LISA typically distributes the number of data streams non-uniformly over the set of active MSs in contrast to the standard hybrid precoding method.}
\revised{The larger gap between all curves and the channel capacity is due to the limited number of RF chains ($N_\text{RF}=8$) at the BS, which limits the maximal slope of the rate curves at high SNR (degrees of freedom), in contrast to the increased degrees of freedom in the unrestricted case of the capacity curve ($\min(N_\text{BS},K\times N_\text{MS},K\times L)=16$).}

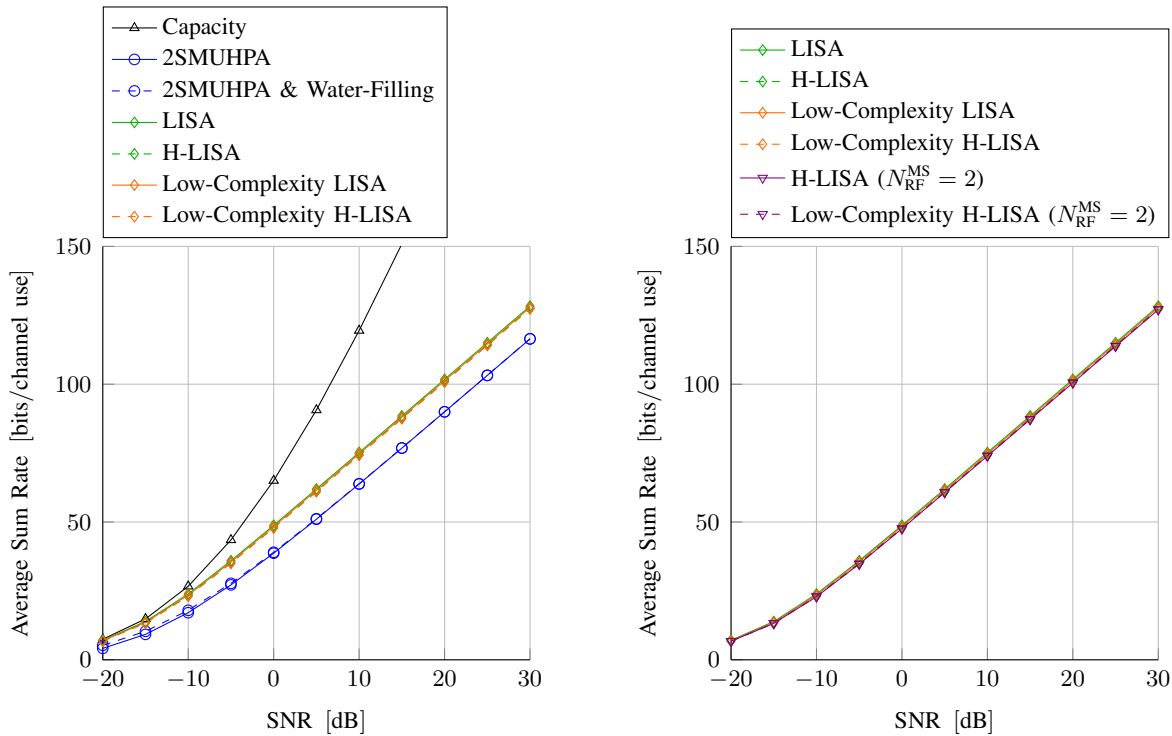
\begin{figure}[h!]
\centering
\subfloat{
\input{Sum_Rate_vs_SNR_GeometricUniform_BSUPA_MSUPA_L3_K8_N_BS64_N_MS16_2017-05-09-02-25-36.tikz}}
\qquad
\subfloat{
\input{Sum_Rate_vs_SNR_GeometricUniform_BSUPA_MSUPA_L3_K8_N_BS64_N_MS16_N_RF_MS2_2017-05-09-02-32-48.tikz}
}
\vspace{-4mm}
\caption{\revised{Average Sum Rate vs. SNR for $N_\text{BS}=64$, $N_\text{MS}=16$, $N_\text{RF}=8$, $K=8$, $L=3$.}}
\label{fig:Simulation_3}
\end{figure}

\revised{A drastic increase of the number of antenna elements at the MS is eventually presented in Fig.~\ref{fig:Simulation_3}, where the average sum rate is plotted over the SNR for MSs equipped with $4\times4$ UPAs ($N_\text{MS}=16$) taking into account a fully digital processing architecture and $L=3$. There is again a significant performance gap between the two versions of the 2SMUHPA and the four versions of LISA. 
Furthermore, it is noteworthy that the versions of H-LISA for analog processing at the MSs with $N_\text{RF}^\text{MS}=2$ RF chains perform only slightly worse than their counterparts for the fully digital processing at the MSs.} At an SNR of $0$ dB, the 2SMUHPA with and without waterfilling achieves approximately $39$ bits per channel use, H-LISA, its low-complexity version, their versions for analog processing at the MSs as well as the low-complexity version of LISA approximately $48$ bits per channel use and LISA approximately $49$ bits per channel use while the capacity is approximately $65$ bits per channel use. \revised{The gap between all curves and the channel capacity at high SNR is now huge, since the degrees of freedom in the unrestricted case of the capacity curve have dramatically grown ($\min(N_\text{BS},K\times N_\text{MS},K\times L)=24$), whereas the number of RF chains is unchanged.}

\section{Conclusion}\label{sec7}

In this paper, we have proposed new multiuser hybrid precoding algorithms for mmWave communications, where a part of the signal processing at the BS serving multiple users is performed in the analog domain by phase shifters and the other part in the digital domain. They are based on the algorithm LISA developed for the traditional fully digital precoding. LISA successively allocates data streams to the MSs, and determines the precoders and equalizers for those data streams, which circumvents the high computational complexity of the direct sum rate maximization. The two stages taken by LISA for suppressing the interstream interference perfectly match the hybrid architecture since it naturally leads to a precoder of a factored form. Using a simple approximation of the first factor for suppressing a part of the interference by a matrix with constant-modulus entries, which can be implemented by phase shifters in the analog domain, in the first step and adapting the second factor for suppressing the remaining interference, which can be implemented in the digital domain, in the second step, we have mapped the two stages of the interference suppression in LISA to the analog and digital domain. This results in a new algorithm for multiuser hybrid precoding, namely, H-LISA. Since this algorithm still has a considerable computational complexity due to the computation of singular values and vectors inherited from LISA, we have developed a low-complexity version of both LISA and H-LISA by exploiting the special structure of the channel matrices for mmWave channels according to the adopted geometric channel model, which does not require the computation of singular values and vectors anymore.

The presented numerical results obtained by simulations demonstrate that the simple approximation by the matrix with constant-modulus entries and the complexity reduction lead only to a slight performance degradation such that both H-LISA and its low-complexity version show excellent performance close to the fully digital version. In view of all numerical results, it can be concluded that the proposed algorithms based on LISA are promising methods for hybrid precoding in multiuser mmWave communication systems. 

\revised{A couple of aspects have not been taken into account in this work and remain subject to further research. First of all, the discussed methods would appropriate channel estimation techniques to acquire the required channel state information (CSI) at the BS and the MSs, which is not discussed in this manuscript.
Futhermore, the proposed solutions in this work are purely based on a single carrier perspective. The extension to wideband systems, e.g., multicarrier systems, remains subject to future research, since simple extensions of narrowband hybrid beamforming solutions are generally not trivial, due to the fixed nature of the analog beamformer for the entire bandwidth.}

\section*{Acknowledgement}
\small{We thank our colleagues from Huawei Technologies D{\"u}sseldorf
GmbH (German Research Center, Munich) and Shanghai Huawei Technologies Co. Ltd., who provided insight and expertise that greatly assisted the research. We especially thank Nikola Vucic, Mario Castaneda, Ronald B{\"o}hnke, Wen Xu, Youtuan Zhu and Long Qin.}

\ifCLASSOPTIONcaptionsoff
  \newpage
\fi

\bibliographystyle{IEEEtran}
\vspace{10pt}
\bibliography{LISA_for_Hybrid_Precoding_2016}

\end{document}

%% file: Sum_Rate_vs_SNR_GeometricUniform_BSUPA_MSULA_L1_K8_N_BS64_N_MS1_2017-05-08-22-08-44.tikz
%
%
\begin{tikzpicture}

\begin{axis}[%
width=0.8\figurewidth,
height=\figureheight,
at={(0\figurewidth,0\figureheight)},
scale only axis,
xmin=-20,
xmax=30,
xlabel={$\text{SNR}~\left[\text{dB}\right]$},
xmajorgrids,
ymin=0,
ymax=100,
ylabel={$\text{Average Sum Rate}~\left[\text{bits}/\text{channel use}\right]$},
ymajorgrids,
axis background/.style={fill=white},
axis x line*=bottom,
axis y line*=left,
legend style={legend cell align=left,align=left,draw=black,at={(0.02,0.78)},anchor=south west}
]
\addplot [color=black,solid,mark=triangle,mark options={solid}]
  table[row sep=crcr]{%
-20	1.53219290880281\\
-15	3.42210163813428\\
-10	6.91708092721467\\
-5	12.5621010788517\\
0	20.5000462022159\\
5	30.3816393626634\\
10	41.6294480008335\\
15	53.7539960709048\\
20	66.4012037402719\\
25	79.3434478792499\\
30	92.4466548322629\\
};
\addlegendentry{Capacity};

\addplot [color=blue,solid,mark=o,mark options={solid}]
  table[row sep=crcr]{%
-20	0.587274373463102\\
-15	1.6678230045006\\
-10	4.19474181091867\\
-5	8.96732747893572\\
0	16.2668694387601\\
5	25.7346564160766\\
10	36.7439240145172\\
15	48.7331249375203\\
20	61.3037099178061\\
25	74.2024048231735\\
30	87.2807182565122\\
};
\addlegendentry{2SMUHPA};

\addplot [color=blue,dashed,mark=o,mark options={solid}]
  table[row sep=crcr]{%
-20	1.25864888022817\\
-15	2.81613866072378\\
-10	5.72432467130936\\
-5	10.5369802943394\\
0	17.5534165801225\\
5	26.6277699240903\\
10	37.2843801086298\\
15	49.0374348859789\\
20	61.4713779592314\\
25	74.2953198285935\\
30	87.3306357485634\\
};
\addlegendentry{2SMUHPA \& Water-Filling};

\addplot [color=black!30!green,solid,mark=diamond,mark options={solid}]
  table[row sep=crcr]{%
-20	1.52185111068131\\
-15	3.37769760609597\\
-10	6.77895409026765\\
-5	12.2172990557904\\
0	19.8114275939605\\
5	29.2247444345152\\
10	39.9221084154774\\
15	51.4986341299967\\
20	63.605167499205\\
25	76.0313459963432\\
30	88.6947295383294\\
};
\addlegendentry{LISA};

\addplot [color=black!30!green,dashed,mark=diamond,mark options={solid}]
  table[row sep=crcr]{%
-20	1.52109808897146\\
-15	3.37408035333045\\
-10	6.76608953905667\\
-5	12.1809409961584\\
0	19.7312990871548\\
5	29.0714382177347\\
10	39.6797942266737\\
15	51.1474049340879\\
20	63.1703128626984\\
25	75.5223217566312\\
30	88.1076831850811\\
};
\addlegendentry{H-LISA};

\addplot [color=red!20!orange,solid,mark=diamond,mark options={solid}]
  table[row sep=crcr]{%
-20	1.52185111068131\\
-15	3.37769760609597\\
-10	6.77895409026765\\
-5	12.2172990557904\\
0	19.8114275939605\\
5	29.2247444345152\\
10	39.9221084154774\\
15	51.4986341299967\\
20	63.605167499205\\
25	76.0313459963432\\
30	88.6947295383293\\
};
\addlegendentry{Low-Complexity LISA};

\addplot [color=red!20!orange,dashed,mark=diamond,mark options={solid}]
  table[row sep=crcr]{%
-20	1.52109808897146\\
-15	3.37408035333045\\
-10	6.76608953905667\\
-5	12.1809409961584\\
0	19.7312990871548\\
5	29.0714382177347\\
10	39.6797942266737\\
15	51.1474049340879\\
20	63.1703128626984\\
25	75.5223217566312\\
30	88.1076831850812\\
};
\addlegendentry{Low-Complexity H-LISA};

\end{axis}
\end{tikzpicture}%

%% file: Sum_Rate_vs_SNR_GeometricUniform_BSUPA_MSULA_L3_K8_N_BS64_N_MS1_2017-05-08-21-35-08.tikz
%
%
\begin{tikzpicture}

\begin{axis}[%
width=0.8\figurewidth,
height=\figureheight,
at={(0\figurewidth,0\figureheight)},
scale only axis,
xmin=-20,
xmax=30,
xlabel={$\text{SNR}~\left[\text{dB}\right]$},
xmajorgrids,
ymin=0,
ymax=100,
ylabel={$\text{Average Sum Rate}~\left[\text{bits}/\text{channel use}\right]$},
ymajorgrids,
axis background/.style={fill=white},
axis x line*=bottom,
axis y line*=left,
legend style={legend cell align=left,align=left,draw=black,at={(0.02,0.78)},anchor=south west}
]
\addplot [color=black,solid,mark=triangle,mark options={solid}]
  table[row sep=crcr]{%
-20	1.2855421699993\\
-15	3.08068031963034\\
-10	6.7176598016407\\
-5	13.0554326752534\\
0	22.2979033204232\\
5	33.7175469424444\\
10	46.289450538133\\
15	59.3327111658672\\
20	72.540996475076\\
25	85.8033645051345\\
30	99.0830390442559\\
};
\addlegendentry{Capacity};

\addplot [color=blue,solid,mark=o,mark options={solid}]
  table[row sep=crcr]{%
-20	0.284778319884994\\
-15	0.857623616299828\\
-10	2.39414007096988\\
-5	5.8329641780078\\
0	11.9310795004597\\
5	20.6462604750773\\
10	31.3256012009371\\
15	43.2496955752221\\
20	55.879650522751\\
25	68.8731749337134\\
30	82.038221707781\\
};
\addlegendentry{2SMUHPA};

\addplot [color=blue,dashed,mark=o,mark options={solid}]
  table[row sep=crcr]{%
-20	0.59845467450864\\
-15	1.48898622338612\\
-10	3.38218874086012\\
-5	6.97950195558718\\
0	12.9255463143498\\
5	21.3229599480013\\
10	31.6992361270526\\
15	43.4252060407319\\
20	55.9509808779424\\
25	68.8980184194524\\
30	82.0451039285047\\
};
\addlegendentry{2SMUHPA \& Water-Filling};

\addplot [color=black!30!green,solid,mark=diamond,mark options={solid}]
  table[row sep=crcr]{%
-20	1.25734517837494\\
-15	2.94886611788099\\
-10	6.28082674112182\\
-5	12.0272262398648\\
0	20.5759196623719\\
5	31.5082716352134\\
10	43.8674732694099\\
15	56.8467290248176\\
20	70.043726264596\\
25	83.305732334684\\
30	96.5854028175512\\
};
\addlegendentry{LISA};

\addplot [color=black!30!green,dashed,mark=diamond,mark options={solid}]
  table[row sep=crcr]{%
-20	1.12526555901203\\
-15	2.66532213403922\\
-10	5.74644926794035\\
-5	11.1477913150847\\
0	19.3271262235697\\
5	29.9702603211949\\
10	42.1689708340641\\
15	55.0849290576728\\
20	68.2621888515771\\
25	81.5193839371579\\
30	94.7976083997794\\
};
\addlegendentry{H-LISA};

\addplot [color=red!20!orange,solid,mark=diamond,mark options={solid}]
  table[row sep=crcr]{%
-20	1.22865470073765\\
-15	2.89121990945843\\
-10	6.18165470239858\\
-5	11.9121612719372\\
0	20.4897263597101\\
5	31.4636211873761\\
10	43.852870675899\\
15	56.8425986972688\\
20	70.0437262645961\\
25	83.3057323346839\\
30	96.5854028175511\\
};
\addlegendentry{Low-Complexity LISA};

\addplot [color=red!20!orange,dashed,mark=diamond,mark options={solid}]
  table[row sep=crcr]{%
-20	1.10234429213991\\
-15	2.61986880574716\\
-10	5.66738541318977\\
-5	11.0495581347999\\
0	19.2496731020448\\
5	29.9340188016884\\
10	42.1626382034809\\
15	55.0887548633538\\
20	68.2722195595642\\
25	81.5294736766126\\
30	94.8077150048394\\
};
\addlegendentry{Low-Complexity H-LISA};

\end{axis}
\end{tikzpicture}%

%% file: Channel_Gains_vs_SNR_GeometricUniform_BSUPA_MSULA_L3_K8_N_BS64_N_MS1_2016-05-07-11-01-20.tikz
%
%
\begin{tikzpicture}

\begin{axis}[%
width=0.951\figurewidth,
height=\figureheight,
at={(0\figurewidth,0\figureheight)},
scale only axis,
xmin=0,
xmax=18,
xlabel={Channel Gain},
xmajorgrids,
ymode=log,
ymin=1,
ymax=10000,
yminorticks=true,
ylabel={Frequency},
ymajorgrids,
yminorgrids,
axis background/.style={fill=white},
axis x line*=bottom,
axis y line*=left,
legend style={legend cell align=left,align=left,draw=white!15!black}
]
\addplot[fill=blue,fill opacity=0.6,draw=black,ybar interval,area legend] plot table[row sep=crcr] {%
x	y\\
0	501\\
0.75	803\\
1.5	787\\
2.25	897\\
3	967\\
3.75	982\\
4.5	906\\
5.25	731\\
6	560\\
6.75	353\\
7.5	240\\
8.25	122\\
9	86\\
9.75	34\\
10.5	13\\
11.25	11\\
12	3\\
12.75	3\\
13.5	0\\
14.25	1\\
15	1\\
};
\addlegendentry{2SMUHPA};


\addplot[fill=black!30!green,fill opacity=0.6,draw=black,ybar interval,area legend] plot table[row sep=crcr] {%
x	y\\
2.25	1\\
3	86\\
3.75	396\\
4.5	796\\
5.25	1051\\
6	1067\\
6.75	980\\
7.5	849\\
8.25	669\\
9	409\\
9.75	279\\
10.5	176\\
11.25	103\\
12	51\\
12.75	35\\
13.5	16\\
14.25	6\\
15	3\\
15.75	1\\
16.5	1\\
};
\addlegendentry{LISA};

\addplot[fill=red,fill opacity=0.6,draw=black,ybar interval,area legend] plot table[row sep=crcr] {%
x	y\\
2.25	8\\
3	188\\
3.75	668\\
4.5	1015\\
5.25	1134\\
6	1076\\
6.75	943\\
7.5	752\\
8.25	463\\
9	305\\
9.75	198\\
10.5	112\\
11.25	51\\
12	37\\
12.75	12\\
13.5	8\\
14.25	1\\
15	3\\
15.75	3\\
};
\addlegendentry{H-LISA};

%

\end{axis}
\end{tikzpicture}%

%% file: Sum_Rate_vs_SNR_GeometricUniform_BSUPA_MSULA_L3_K8_N_BS64_N_MS2_N_RF_MS2_2017-05-08-22-17-53.tikz
%
%
\begin{tikzpicture}

\begin{axis}[%
width=\figurewidth,
height=\figureheight,
at={(0\figurewidth,0\figureheight)},
scale only axis,
xmin=-20,
xmax=30,
xlabel={$\text{SNR}~\left[\text{dB}\right]$},
xmajorgrids,
ymin=0,
ymax=120,
ylabel={$\text{Average Sum Rate}~\left[\text{bits}/\text{channel use}\right]$},
ymajorgrids,
axis background/.style={fill=white},
axis x line*=bottom,
axis y line*=left,
legend style={legend cell align=left,align=left,draw=black,at={(1.02,0.98)},anchor=north west}
]
\addplot [color=black,solid,mark=triangle,mark options={solid}]
  table[row sep=crcr]{%
-20	1.97196948309093\\
-15	4.52330540270021\\
-10	9.40973945318477\\
-5	17.5268618294752\\
0	29.3481789748855\\
5	44.9202573707858\\
10	63.8286265531976\\
15	85.4104476085234\\
20	108.944387600008\\
25	133.779746232027\\
30	159.413617258669\\
};
\addlegendentry{Capacity};

\addplot [color=blue,solid,mark=o,mark options={solid}]
  table[row sep=crcr]{%
-20	0.601589172008479\\
-15	1.73730517971017\\
-10	4.46885529449489\\
-5	9.71398872665793\\
0	17.6921501363957\\
5	27.8461592267038\\
10	39.4225037770009\\
15	51.8324182103589\\
20	64.6948495680708\\
25	77.785878340388\\
30	90.9841173940699\\
};
\addlegendentry{2SMUHPA};

\addplot [color=blue,dashed,mark=o,mark options={solid}]
  table[row sep=crcr]{%
-20	1.10770315564115\\
-15	2.60721630773948\\
-10	5.59290582424445\\
-5	10.7905710145661\\
0	18.4966178985432\\
5	28.3423915299453\\
10	39.6796334684506\\
15	51.9491045857332\\
20	64.7429472509003\\
25	77.8070416988094\\
30	90.9944571088757\\
};
\addlegendentry{2SMUHPA \& Water-Filling};

\addplot [color=red,solid,mark=square,mark options={solid}]
  table[row sep=crcr]{%
-20	1.56730084109382\\
-15	3.37326333678774\\
-10	6.53716132610525\\
-5	11.4700186755131\\
0	18.4296578543883\\
5	27.4096548779407\\
10	38.0641632785736\\
15	49.8992705569232\\
20	62.4649164314531\\
25	75.4098857869857\\
30	88.5479266423244\\
};
\addlegendentry{BD};

\addplot [color=black!30!green,solid,mark=diamond,mark options={solid}]
  table[row sep=crcr]{%
-20	1.91893592062412\\
-15	4.3127119684299\\
-10	8.78963521308536\\
-5	16.1293403963489\\
0	26.4299831281599\\
5	38.5907457870049\\
10	51.5199343650442\\
15	64.6960535495626\\
20	77.9481176576039\\
25	91.224531430473\\
30	104.508668303913\\
};
\addlegendentry{LISA};

\addplot [color=black!30!green,dashed,mark=diamond,mark options={solid}]
  table[row sep=crcr]{%
-20	1.75512801488162\\
-15	3.98357409586431\\
-10	8.20341502644665\\
-5	15.215312915488\\
0	25.2088871553219\\
5	37.2029251479928\\
10	50.0728630687277\\
15	63.2307669323691\\
20	76.4776778802803\\
25	89.7524539567357\\
30	103.036072127675\\
};
\addlegendentry{H-LISA};

\addplot [color=red!20!orange,solid,mark=diamond,mark options={solid}]
  table[row sep=crcr]{%
-20	1.84075592848097\\
-15	4.15604500891817\\
-10	8.52951414756737\\
-5	15.7516280079541\\
0	25.9597534593759\\
5	38.0720402184628\\
10	50.9660450101856\\
15	64.1296791087212\\
20	77.3794200935339\\
25	90.6550954678304\\
30	103.938998460558\\
};
\addlegendentry{Low-Complexity LISA};

\addplot [color=red!20!orange,dashed,mark=diamond,mark options={solid}]
  table[row sep=crcr]{%
-20	1.70606280405506\\
-15	3.88532777873858\\
-10	8.04606674169062\\
-5	14.992381973006\\
0	24.9393994641419\\
5	36.9060268623818\\
10	49.7432609778358\\
15	62.8904193843231\\
20	76.1356145100048\\
25	89.4098445930738\\
30	102.69328976289\\
};
\addlegendentry{Low-Complexity H-LISA};

%

\end{axis}
\end{tikzpicture}%

%% file: Sum_Rate_vs_SNR_GeometricUniform_BSUPA_MSUPA_L3_K8_N_BS64_N_MS16_2017-05-09-02-25-36.tikz
%
%
\begin{tikzpicture}

\begin{axis}[%
width=0.8\figurewidth,
height=\figureheight,
at={(0\figurewidth,0\figureheight)},
scale only axis,
xmin=-20,
xmax=30,
xlabel={$\text{SNR}~\left[\text{dB}\right]$},
xmajorgrids,
ymin=0,
ymax=150,
ylabel={$\text{Average Sum Rate}~\left[\text{bits}/\text{channel use}\right]$},
ymajorgrids,
axis background/.style={fill=white},
axis x line*=bottom,
axis y line*=left,
legend style={legend cell align=left,align=left,draw=black,at={(0,1.02)},anchor=south west}
]
\addplot [color=black,solid,mark=triangle,mark options={solid}]
  table[row sep=crcr]{%
-20	7.45328580508084\\
-15	14.8375942430186\\
-10	26.6754003730541\\
-5	43.4844628317926\\
0	65.034723117971\\
5	90.6289362900925\\
10	119.468687142973\\
15	150.866417448286\\
20	184.217887225144\\
25	219.096911905473\\
30	255.169121118832\\
};
\addlegendentry{Capacity};

\addplot [color=blue,solid,mark=o,mark options={solid}]
  table[row sep=crcr]{%
-20	4.14831834760265\\
-15	9.21924520140275\\
-10	17.085187965772\\
-5	27.1720140650319\\
0	38.6778405141633\\
5	51.0031154401888\\
10	63.7830347587236\\
15	76.8106781532852\\
20	89.9700428521892\\
25	103.196383931614\\
30	116.455677344478\\
};
\addlegendentry{2SMUHPA};

\addplot [color=blue,dashed,mark=o,mark options={solid}]
  table[row sep=crcr]{%
-20	5.25056067016486\\
-15	10.3265604948775\\
-10	17.9548986288765\\
-5	27.7494457860747\\
0	39.0142792378078\\
5	51.1863566899481\\
10	63.8742815465858\\
15	76.8523178425867\\
20	89.98919898925\\
25	103.205054361181\\
30	116.458912023053\\
};
\addlegendentry{2SMUHPA \& Water-Filling};

\addplot [color=black!30!green,solid,mark=diamond,mark options={solid}]
  table[row sep=crcr]{%
-20	7.12035817163113\\
-15	13.8815319623536\\
-10	23.8985942865316\\
-5	35.9390356003856\\
0	48.8098400469016\\
5	61.9636895622681\\
10	75.2087206116947\\
15	88.4829005000899\\
20	101.766329968722\\
25	115.052687616312\\
30	128.339971556324\\
};
\addlegendentry{LISA};

\addplot [color=black!30!green,dashed,mark=diamond,mark options={solid}]
  table[row sep=crcr]{%
-20	6.86048048860675\\
-15	13.4272529323396\\
-10	23.2435011366092\\
-5	35.176015236407\\
0	48.0080116167186\\
5	61.1510307545553\\
10	74.392576130732\\
15	87.6656475450855\\
20	100.948725858611\\
25	114.234972398591\\
30	127.522221197027\\
};
\addlegendentry{H-LISA};

\addplot [color=red!20!orange,solid,mark=diamond,mark options={solid}]
  table[row sep=crcr]{%
-20	6.90889113966011\\
-15	13.5480263274369\\
-10	23.4817114264347\\
-5	35.4726775587514\\
0	48.3224161007932\\
5	61.4711755626794\\
10	74.7143487857641\\
15	87.9879378308218\\
20	101.271180121409\\
25	114.557478544333\\
30	127.844743752483\\
};
\addlegendentry{Low-Complexity LISA};

\addplot [color=red!20!orange,dashed,mark=diamond,mark options={solid}]
  table[row sep=crcr]{%
-20	6.7642726829509\\
-15	13.2984648822863\\
-10	23.1066704077205\\
-5	35.0218969118742\\
0	47.8493409811498\\
5	60.9889481555105\\
10	74.2300195892651\\
15	87.5029400399463\\
20	100.785970504623\\
25	114.072201902444\\
30	127.359445911406\\
};
\addlegendentry{Low-Complexity H-LISA};

\end{axis}
\end{tikzpicture}%

%% file: Sum_Rate_vs_SNR_GeometricUniform_BSUPA_MSUPA_L3_K8_N_BS64_N_MS16_N_RF_MS2_2017-05-09-02-32-48.tikz
%
%
\begin{tikzpicture}

\begin{axis}[%
width=0.8\figurewidth,
height=\figureheight,
at={(0\figurewidth,0\figureheight)},
scale only axis,
xmin=-20,
xmax=30,
xlabel={$\text{SNR}~\left[\text{dB}\right]$},
xmajorgrids,
ymin=0,
ymax=150,
ylabel={$\text{Average Sum Rate}~\left[\text{bits}/\text{channel use}\right]$},
ymajorgrids,
axis background/.style={fill=white},
axis x line*=bottom,
axis y line*=left,
legend style={legend cell align=left,align=left,draw=black,at={(0,1.02)},anchor=south west}
]
%
%

\addplot [color=black!30!green,solid,mark=diamond,mark options={solid}]
  table[row sep=crcr]{%
-20	7.12035817163113\\
-15	13.8815319623536\\
-10	23.8985942865316\\
-5	35.9390356003856\\
0	48.8098400469016\\
5	61.9636895622681\\
10	75.2087206116947\\
15	88.4829005000899\\
20	101.766329968722\\
25	115.052687616312\\
30	128.339971556324\\
};
\addlegendentry{LISA};

\addplot [color=black!30!green,dashed,mark=diamond,mark options={solid}]
  table[row sep=crcr]{%
-20	6.86048048860675\\
-15	13.4272529323396\\
-10	23.2435011366092\\
-5	35.176015236407\\
0	48.0080116167186\\
5	61.1510307545553\\
10	74.392576130732\\
15	87.6656475450855\\
20	100.948725858611\\
25	114.234972398591\\
30	127.522221197027\\
};
\addlegendentry{H-LISA};

\addplot [color=red!20!orange,solid,mark=diamond,mark options={solid}]
  table[row sep=crcr]{%
-20	6.90889113966011\\
-15	13.5480263274369\\
-10	23.4817114264347\\
-5	35.4726775587514\\
0	48.3224161007932\\
5	61.4711755626794\\
10	74.7143487857641\\
15	87.9879378308218\\
20	101.271180121409\\
25	114.557478544333\\
30	127.844743752483\\
};
\addlegendentry{Low-Complexity LISA};

\addplot [color=red!20!orange,dashed,mark=diamond,mark options={solid}]
  table[row sep=crcr]{%
-20	6.7642726829509\\
-15	13.2984648822863\\
-10	23.1066704077205\\
-5	35.0218969118742\\
0	47.8493409811498\\
5	60.9889481555105\\
10	74.2300195892651\\
15	87.5029400399463\\
20	100.785970504623\\
25	114.072201902444\\
30	127.359445911406\\
};
\addlegendentry{Low-Complexity H-LISA};

\addplot [color=violet,solid,mark=triangle,mark options={solid,rotate=180}]
table[row sep=crcr]{%
	-20	6.75372313176931\\
	-15	13.2123319433188\\
	-10	22.8891255118446\\
	-5	34.7493644215046\\
	0	47.5577063707993\\
	5	60.6933023327048\\
	10	73.9326113429935\\
	15	87.2049708748602\\
	20	100.487823599608\\
	25	113.773998754861\\
	30	127.061224974731\\
};
\addlegendentry{H-LISA ($N_\text{RF}^\text{MS}=2$)};

\addplot [color=violet,dashed,mark=triangle,mark options={solid,rotate=180}]
  table[row sep=crcr]{%
-20	6.74717589896852\\
-15	13.234444349018\\
-10	22.9721469634362\\
-5	34.8513915810169\\
0	47.6649109038565\\
5	60.8004306512298\\
10	74.040533070185\\
15	87.3131450304262\\
20	100.59607773414\\
25	113.882278196406\\
30	127.169512420602\\
};
\addlegendentry{Low-Complexity H-LISA ($N_\text{RF}^\text{MS}=2$)};

\end{axis}
\end{tikzpicture}%